\documentclass[5p,times]{elsarticle} %

\usepackage{xspace} %
\usepackage{amsmath} %
\usepackage{hyperref}
\usepackage[capitalise,nameinlink,noabbrev]{cleveref} %
\usepackage{siunitx} %
\usepackage{csquotes} %
\usepackage{booktabs} %
\usepackage{float} %
\usepackage{hyphenat} %
\usepackage{caption} %
\usepackage{microtype} %
\usepackage{pgfplots} %
\usepackage{mhchem} %
\usepackage{xltabular} %
\usepackage[abbreviations,acronym,prefix,shortcuts=ac,section=subsection,toc=false,symbols]{glossaries-extra}
\usepackage{xurl} %
\usepackage[showseconds=false,showzone=false,showzoneminutes=false]{datetime2} %
\usepackage{xcolor}
\usepackage[normalem]{ulem} %

\hypersetup{colorlinks=true} %
\AtBeginDocument{\hypersetup{linkcolor=cyan,citecolor=cyan,urlcolor=cyan}} %

\crefformat{equation}{(#2#1#3)}
\crefrangeformat{equation}{(#3#1#4)--(#5#2#6)}
\crefmultiformat{equation}{(#2#1#3)}{, (#2#1#3)}{, (#2#1#3)}{ and~(#2#1#3)}
\crefrangemultiformat{equation}{(#3#1#4)--(#5#2#6)}{, (#3#1#4)--(#5#2#6)}{, (#3#1#4)--(#5#2#6)}{ and~(#3#1#4)--(#5#2#6)}

\Crefformat{equation}{Equation~(#2#1#3)}
\Crefrangeformat{equation}{Equations~(#3#1#4)--(#5#2#6)}
\Crefmultiformat{equation}{Equations~(#2#1#3)}{, (#2#1#3)}{, (#2#1#3)}{ and~(#2#1#3)}
\Crefrangemultiformat{equation}{Equations~(#3#1#4)--(#5#2#6)}{, (#3#1#4)--(#5#2#6)}{, (#3#1#4)--(#5#2#6)}{ and~(#3#1#4)--(#5#2#6)}

\def\appendixname{} %

\makenoidxglossaries

\setabbreviationstyle[\acronymtype]{long-short}

\glsaddstoragekey{unit}{}{\glsunit}
\glsaddstoragekey{magnitude}{}{\glsmagnitude}

\glssetnoexpandfield{unit}
\glssetnoexpandfield{magnitude}

\newglossary*{constants}{Constants}

\glsdisablehyper %

\pgfplotsset{compat=1.17}
\usepgfplotslibrary{groupplots}
\usepgfplotslibrary{units}
\usepgfplotslibrary{patchplots} %
\usetikzlibrary{matrix} %
\usetikzlibrary{shapes}
\usetikzlibrary{patterns}
\usetikzlibrary{arrows.meta}
\usetikzlibrary{positioning}
\usetikzlibrary{backgrounds}
\usetikzlibrary{shadows}
\usetikzlibrary{arrows}
\usetikzlibrary{calc}
\usetikzlibrary{automata} %

\pgfplotsset{unit markings=slash space} %

\tikzset{>=latex[round]}
\tikzstyle{*->} = [{Circle[length=3pt]}->, shorten <=-1.5pt]
\tikzstyle{<-*} = [<-{Circle[length=3pt]}, shorten >=-1.5pt]

\pgfplotsset{
	compat=1.11,
	legend image code/.code={
		\draw[mark repeat=2,mark phase=2]
		plot coordinates {
			(0cm,0cm)
			(0.15cm,0cm)        %
			(0.3cm,0cm)         %
		};%
	}
}

\pgfplotsset{
	/pgfplots/layers/niceLayers/.define layer set={
		axis background,
		axis grid,
		main,
		axis ticks,
		axis lines,
		axis tick labels,
		axis descriptions,
		axis foreground
	}{/pgfplots/layers/standard}
}

\pgfplotsset{
	every axis/.append style={
		set layers=niceLayers,
		tick label style={font=\scriptsize},
		clip marker paths=true,
		line width=1pt,
		line cap=round,
		line join=round,
		tick style={semithick, color=black},
		legend style={
			/tikz/every even column/.append style={column sep=2mm},
			cells={anchor=west}, %
		},
		xmajorgrids,
		ymajorgrids,
	}
}

\definecolor{color1}{HTML}{648fff} %
\definecolor{color2}{HTML}{dc267f} %
\definecolor{color3}{HTML}{ffb000} %
\definecolor{color4}{HTML}{785ef0} %
\definecolor{color5}{HTML}{fe6100} %

\definecolor{color6}{HTML}{13AB8E} %

\colorlet{heatingColor}{color2}
\colorlet{coolingColor}{color1}

\colorlet{pmv025Color}{color1}
\colorlet{pmv1Color}{color2}
\colorlet{gohlichColor}{color3}
\colorlet{hambraeusColor}{color4}

\pgfplotsset{
    passiveStyle/.style={
		only marks, 
		mark=*, 
		mark options={solid, white!80!black},
		mark size=0.5pt,
	},
	coolingStyle/.style={
		only marks, 
		mark=*, 
		mark options={solid, coolingColor!50!white},
		mark size=0.3pt,
	},
	heatingStyle/.style={
		only marks, 
		mark=*, 
		mark options={solid, heatingColor!50!white},
		mark size=0.3pt,
	},
	passiveStyleLegendEntry/.style={
		passiveStyle,
		mark size=3pt,
	},
	coolingStyleLegendEntry/.style={
		coolingStyle,
		mark size=3pt,
	},
	heatingStyleLegendEntry/.style={
		heatingStyle,
		mark size=3pt,
	},
	setpointStyle/.style={
		very thick, 
		dashed,
	},
    dynamicStyle/.style={
	},
    staticStyle/.style={
		dashed, 
		line width=2pt, 
		const plot, 
		jump mark left, 
		forget plot,
		line cap = butt, %
	},
	refStyle/.style={
		dotted,
		thick,
		unbounded coords=jump,
		black,
	},
}

\tikzset{
	pmvWindowMarkingRectangleStyle/.style={
		rounded corners=3pt,
		draw,
		dotted,
		thick,
		opacity=0.3
	},
	pmvWindowMarkingLabelStyle/.style={
		align=center,
		font=\footnotesize,
		color=white!50!black
	},
}

\def\legendoffset{0.075cm}
\newcommand{\legendNorthEast}{
at={($(rel axis cs:1,1) - (\legendoffset,\legendoffset)$)},
anchor=north east,
}
\newcommand{\legendNorthWest}{
at={($(rel axis cs:0,1) + (\legendoffset,-\legendoffset)$)},
anchor=north west,
}

\pgfkeys{/pgf/number format/hour code/.code={\pgfmathparse{int(#1)}\pgfmathresult:00}}

\newcommand{\timeDerivative}{\frac{\mathrm{d}}{\mathrm{d}t}} %

\newcommand{\PMVRange}[1]{\ensuremath{\PMVvar \in [-#1, +#1]}}
\newcommand{\PMVRangeTime}[1]{\ensuremath{\PMVvar(t) \in [-#1, +#1]}}

\journal{Control Engineering Practice}

\newglossaryentry{Nsamples}{name=\ensuremath{N_\mathrm{samples}},
type=constants,
sort=Nsamples,
description={Number of samples considered},
magnitude=\num{7399},
}
\newcommand{\Nsamples}{\gls{Nsamples}\xspace}

\newglossaryentry{disturbanceVector}{name=\ensuremath{\mathbf{d}},
type=symbols,
sort=d,
description={Disturbance vector},
}
\newcommand{\disturbanceVector}{\gls{disturbanceVector}\xspace}

\newglossaryentry{diffuseIrradiance}{name=\ensuremath{I_\mathrm{dhi}},
type=symbols,
sort=Idhi,
description={\Acs{DHI}},
}
\newcommand{\diffuseIrradiance}{\gls{diffuseIrradiance}\xspace}

\newglossaryentry{beamIrradiance}{name=\ensuremath{I_\mathrm{dni}},
type=symbols,
sort=Idni,
description={\Acs{DNI}},
}
\newcommand{\beamIrradiance}{\gls{beamIrradiance}\xspace}

\newglossaryentry{roofIrradiance}{name=\ensuremath{I_\mathrm{roof}},
type=symbols,
sort=Iroof,
description={Solar irradiance on the cabin roof},
}
\newcommand{\roofIrradiance}{\gls{roofIrradiance}\xspace}

\newglossaryentry{meanWallIrradiance}{name=\ensuremath{I_\mathrm{wall}},
type=symbols,
sort=Iwall,
description={Solar irradiance on the cabin walls},
}
\newcommand{\meanWallIrradiance}{\gls{meanWallIrradiance}\xspace}

\newglossaryentry{Npass}{name=\ensuremath{N_\mathrm{pass}},
type=symbols,
sort=Npass,
description={Number of passengers in the bus},
}
\newcommand{\Npass}{\gls{Npass}\xspace}

\newglossaryentry{QsolarInt}{name=\ensuremath{\dot{Q}_\mathrm{sol,int}},
type=symbols,
sort=Qsolint,
description={Solar heat to bus interior},
}
\newcommand{\QsolarInt}{\gls{QsolarInt}\xspace}

\newglossaryentry{QsolarShellInside}{name=\ensuremath{\dot{Q}_\mathrm{sol,s,i}},
type=symbols,
sort=Qsolsi,
description={Solar heat to shell inside},
}
\newcommand{\QsolarShellInside}{\gls{QsolarShellInside}\xspace}

\newglossaryentry{QsolarShellOutside}{name=\ensuremath{\dot{Q}_\mathrm{sol,s,o}},
type=symbols,
sort=Qsolso,
description={Solar heat to shell outside},
}
\newcommand{\QsolarShellOutside}{\gls{QsolarShellOutside}\xspace}

\newglossaryentry{timeVar}{name=\ensuremath{t},
type=symbols,
sort=t,
description={Absolute local date and time},
}
\newcommand{\timeVar}{\gls{timeVar}\xspace}

\newglossaryentry{Tamb}{name=\ensuremath{T_\mathrm{\infty}},
type=symbols,
sort=Tamb,
description={Ambient temperature},
}
\newcommand{\Tamb}{\gls{Tamb}\xspace}

\newglossaryentry{solarAltitude}{name=\ensuremath{\beta},
type=symbols,
sort=Zbeta,
description={Solar altitude angle},
}
\newcommand{\solarAltitude}{\gls{solarAltitude}\xspace}

\newglossaryentry{solarAzimuth}{name=\ensuremath{\phi},
type=symbols,
sort=Zphi,
description={Solar azimuth angle},
}

\newglossaryentry{surfaceAzimuth}{name=\ensuremath{\psi},
type=symbols,
sort=Zpsi,
description={Surface normal azimuth angle},
}

\newglossaryentry{doorOpenFraction}{name=\ensuremath{\zeta_\mathrm{door}},
type=symbols,
sort=Zzetadoor,
description={Door open fraction},
}
\newcommand{\doorOpenFraction}{\gls{doorOpenFraction}\xspace}

\newglossaryentry{Ai}{name=\ensuremath{A_i},
type=constants,
sort=Ai,
description={Area of passenger surface $i$},
unit=\unit{\square\meter},
}
\newcommand{\Ai}{\gls{Ai}\xspace}

\newglossaryentry{Aint}{name=\ensuremath{A_\mathrm{int}},
type=constants,
sort=Aint,
description={Surface area of the interior},
magnitude=\num{20},
unit=\unit{\square\meter},
}
\newcommand{\Aint}{\gls{Aint}\xspace}

\newglossaryentry{ARH}{name=\ensuremath{A_\mathrm{rh}},
type=constants,
sort=Arh,
description={\Acs{RH} surface area},
magnitude=\num{4},
unit=\unit{\square\meter},
}
\newcommand{\ARH}{\gls{ARH}\xspace}

\newglossaryentry{roofArea}{name=\ensuremath{A_\mathrm{roof}},
type=constants,
sort=Aroof,
description={Roof area},
magnitude=\num{48.6},
unit=\unit{\square\meter},
}
\newcommand{\roofArea}{\gls{roofArea}\xspace}

\newglossaryentry{Ahull}{name=\ensuremath{A_\mathrm{s}},
type=constants,
sort=As,
description={Total cabin shell area},
magnitude=\num{199.5},
unit=\unit{\square\meter},
}
\newcommand{\Ahull}{\gls{Ahull}\xspace}

\newglossaryentry{wallArea}{name=\ensuremath{A_\mathrm{wall}},
type=constants,
sort=Awall,
description={Cabin wall area},
magnitude=\num{102.2},
unit=\unit{\square\meter},
}
\newcommand{\wallArea}{\gls{wallArea}\xspace}

\newglossaryentry{heatCapacityCabin}{name=\ensuremath{C_\mathrm{cab}},
type=constants,
sort=Ccab,
description={Heat capacity cabin air},
magnitude=\num{146.6},
unit=\unit{\kilo\joule\per\kelvin},
}
\newcommand{\heatCapacityCabin}{\gls{heatCapacityCabin}\xspace}

\newglossaryentry{dischargeCoefficient}{name=\ensuremath{C_\mathrm{d}},
type=constants,
sort=Cd,
description={Door model discharge coefficient},
magnitude=\num{0.6},
}
\newcommand{\dischargeCoefficient}{\gls{dischargeCoefficient}\xspace}

\newglossaryentry{heatCapacityInterior}{name=\ensuremath{C_\mathrm{int}},
type=constants,
sort=Cint,
description={Heat capacity interior},
magnitude=\num{78},
unit=\unit{\kilo\joule\per\kelvin},
}
\newcommand{\heatCapacityInterior}{\gls{heatCapacityInterior}\xspace}

\newglossaryentry{airHeatCapacity}{name=\ensuremath{c_\mathrm{p,air}},
type=constants,
sort=cpair,
description={Specific heat capacity of air},
magnitude=\num{1005},
unit=\unit{\joule\per\kilogram\per\kelvin},
}
\newcommand{\airHeatCapacity}{\gls{airHeatCapacity}\xspace}

\newglossaryentry{heatCapacityRH}{name=\ensuremath{C_\mathrm{rh}},
type=constants,
sort=Crh,
description={Heat capacity of \acsp{RH}},
magnitude=\num{4800},
unit=\unit{\joule\per\kelvin},
}
\newcommand{\heatCapacityRH}{\gls{heatCapacityRH}\xspace}

\newglossaryentry{heatCapacityShellInside}{name=\ensuremath{C_\mathrm{s,i}},
type=constants,
sort=Csi,
description={Heat capacity cabin shell inside},
magnitude=\num{856.1},
unit=\unit{\kilo\joule\per\kelvin},
}
\newcommand{\heatCapacityShellInside}{\gls{heatCapacityShellInside}\xspace}

\newglossaryentry{heatCapacityShellOutside}{name=\ensuremath{C_\mathrm{s,o}},
type=constants,
sort=Cso,
description={Heat capacity cabin shell outside},
magnitude=\num{856.1},
unit=\unit{\kilo\joule\per\kelvin},
}
\newcommand{\heatCapacityShellOutside}{\gls{heatCapacityShellOutside}\xspace}

\newglossaryentry{pmvAnnFunction}{name=\ensuremath{f_{\mathrm{ann},T_\mathrm{\infty}}},
type=constants,
sort=fann,
description={\Acs{ANN} function predicting \acs{PMV} (see \cref{eq:pmv-approx})},
unit=\unit{-},
}
\newcommand{\pmvAnnFunction}{\gls{pmvAnnFunction}\xspace}

\newglossaryentry{FIntToShell}{name=\ensuremath{F_\mathrm{int\rightarrow{}s,i}},
type=constants,
sort=Fint2si,
description={View factor interior to shell inside},
magnitude=\num{94},
unit=\unit{\percent},
}
\newcommand{\FIntToShell}{\gls{FIntToShell}\xspace}

\newglossaryentry{FrhToInt}{name=\ensuremath{F_\mathrm{rh\rightarrow{}int}},
type=constants,
sort=Frh2int,
description={View factor \acs{RH} to interior},
magnitude=\num{30},
unit=\unit{\percent},
}
\newcommand{\FrhToInt}{\gls{FrhToInt}\xspace}

\newglossaryentry{FrhToShell}{name=\ensuremath{F_\mathrm{rh\rightarrow{}s,i}},
type=constants,
sort=Frh2si,
description={View factor \acs{RH} to shell inside},
magnitude=\num{70},
unit=\unit{\percent},
}
\newcommand{\FrhToShell}{\gls{FrhToShell}\xspace}

\newglossaryentry{cabinHeight}{name=\ensuremath{h_\mathrm{cab}},
type=constants,
sort=hcab,
description={Cabin height},
magnitude=\num{2.4},
unit=\unit{\meter},
}
\newcommand{\cabinHeight}{\gls{cabinHeight}\xspace}

\newglossaryentry{doorHeight}{name=\ensuremath{h_\mathrm{door}},
type=constants,
sort=hdoor,
description={Height of the doors},
magnitude=\num{1.95},
unit=\unit{\meter},
}
\newcommand{\doorHeight}{\gls{doorHeight}\xspace}

\newglossaryentry{hInside}{name=\ensuremath{h_\mathrm{in}},
type=constants,
sort=hin,
description={Conv.\ heat transfer coeff.\ inside},
magnitude=\num{8.01},
unit=\unit{\watt\per\square\meter\per\kelvin},
}
\newcommand{\hInside}{\gls{hInside}\xspace}

\newglossaryentry{hOutside}{name=\ensuremath{h_\mathrm{out}},
type=constants,
sort=hout,
description={Conv.\ heat transfer coeff.\ outside},
magnitude=\num{20.67},
unit=\unit{\watt\per\square\meter\per\kelvin},
}
\newcommand{\hOutside}{\gls{hOutside}\xspace}

\newglossaryentry{hRh}{name=\ensuremath{h_\mathrm{rh}},
type=constants,
sort=hrh,
description={Conv.\ heat transfer coeff.\ at \acs{RH}},
magnitude=\num{2.1},
unit=\unit{\watt\per\square\meter\per\kelvin},
}
\newcommand{\hRh}{\gls{hRh}\xspace}

\newglossaryentry{conductance}{name=\ensuremath{k_\mathrm{s}},
type=constants,
sort=ks,
description={Heat conduction coeff.\ through shell},
magnitude=\num{6.86},
unit=\unit{\watt\per\square\meter\per\kelvin},
}
\newcommand{\conductance}{\gls{conductance}\xspace}

\newglossaryentry{cabinLength}{name=\ensuremath{l_\mathrm{cab}},
type=constants,
sort=lcab,
description={Cabin length},
magnitude=\num{18.7},
unit=\unit{\meter},
}
\newcommand{\cabinLength}{\gls{cabinLength}\xspace}

\newglossaryentry{airCurtainSpecification}{name=\ensuremath{P_\mathrm{aircurt,0}},
type=constants,
sort=Paircurt0,
description={Specified air curtain power usage},
magnitude=\num{1020},
unit=\unit{\watt},
}
\newcommand{\airCurtainSpecification}{\gls{airCurtainSpecification}\xspace}

\newglossaryentry{metabolicHeat}{name=\ensuremath{\dot{Q}_\mathrm{met}},
type=constants,
sort=Qmet,
description={Metabolic heat rate of one passenger},
magnitude=\num{125.3},
unit=\unit{\watt},
}
\newcommand{\metabolicHeat}{\gls{metabolicHeat}\xspace}

\newglossaryentry{depotTemperature}{name=\ensuremath{T_\mathrm{depot}},
type=constants,
sort=Tdepot,
description={Depot temperature (see \cref{eq:depot-temperature})},
unit=\unit{\kelvin},
}
\newcommand{\depotTemperature}{\gls{depotTemperature}\xspace}

\newglossaryentry{TRHTarget}{name=\ensuremath{T_\mathrm{rh,target}},
type=constants,
sort=Trhtarget,
description={\Acs{RH} target temperature},
magnitude=\num{70},
unit=\unit{\celsius},
}
\newcommand{\TRHTarget}{\gls{TRHTarget}\xspace}

\newglossaryentry{vCab}{name=\ensuremath{v_\mathrm{cab}},
type=constants,
sort=vcab,
description={Draft velocity inside the bus},
magnitude=\num{0.1},
unit=\unit{\meter\per\second},
}
\newcommand{\vCab}{\gls{vCab}\xspace}

\newglossaryentry{cabinVolume}{name=\ensuremath{V_\mathrm{cab}},
type=constants,
sort=Vcab,
description={Cabin air volume},
magnitude=\num{116.7},
unit=\unit{\meter\cubed},
}
\newcommand{\cabinVolume}{\gls{cabinVolume}\xspace}

\newglossaryentry{cabinWidth}{name=\ensuremath{w_\mathrm{cab}},
type=constants,
sort=wcab,
description={Cabin width},
magnitude=\num{2.6},
unit=\unit{\meter},
}
\newcommand{\cabinWidth}{\gls{cabinWidth}\xspace}

\newglossaryentry{totalDoorWidth}{name=\ensuremath{w_\mathrm{door,tot}},
type=constants,
sort=wdoortot,
description={Combined width of the bus doors},
magnitude=\num{4.42},
unit=\unit{\meter},
}
\newcommand{\totalDoorWidth}{\gls{totalDoorWidth}\xspace}

\newglossaryentry{surfaceAbsorptivity}{name=\ensuremath{\alpha_\mathrm{paint}},
type=constants,
sort=Zalphapaint,
description={Solar absorptivity of the bus shell},
magnitude=\num{30},
unit=\unit{\percent},
}
\newcommand{\surfaceAbsorptivity}{\gls{surfaceAbsorptivity}\xspace}

\newglossaryentry{COPvar}{name=\ensuremath{\gamma},
type=constants,
sort=Zgamma,
description={\Acs{COP} (see \cref{fig:model-parameters})},
unit=\unit{-},
}
\newcommand{\COPvar}{\gls{COPvar}\xspace}

\newglossaryentry{COPAC}{name=\ensuremath{\gamma_\mathrm{ac}},
type=constants,
sort=Zgammaac,
description={\Acs{COP} in cooling mode (see \cref{fig:model-parameters})},
unit=\unit{-},
}
\newcommand{\COPAC}{\gls{COPAC}\xspace}

\newglossaryentry{COPHP}{name=\ensuremath{\gamma_\mathrm{hp}},
type=constants,
sort=Zgammahp,
description={\Acs{COP} in heating mode (see \cref{fig:model-parameters})},
unit=\unit{-},
}
\newcommand{\COPHP}{\gls{COPHP}\xspace}

\newglossaryentry{rhCab}{name=\ensuremath{\phi_\mathrm{cab}},
type=constants,
sort=Zphicab,
description={Relative humidity in the cabin},
magnitude=\num{40},
unit=\unit{\percent},
}
\newcommand{\rhCab}{\gls{rhCab}\xspace}

\newglossaryentry{airDensity}{name=\ensuremath{\rho_\mathrm{air}},
type=constants,
sort=Zrhoair,
description={Air density},
magnitude=\num{1.25},
unit=\unit{\kilogram\per\cubic\meter},
}
\newcommand{\airDensity}{\gls{airDensity}\xspace}

\newglossaryentry{timeConstVcc}{name=\ensuremath{\tau_\mathrm{vcc}},
type=constants,
sort=Ztauvcc,
description={Time const. of first order \acs{VCC} dynamics},
magnitude=\num{20},
unit=\unit{\second},
}
\newcommand{\timeConstVcc}{\gls{timeConstVcc}\xspace}

\newglossaryentry{windowTransmissivity}{name=\ensuremath{\tau_\mathrm{win}},
type=constants,
sort=Ztauwin,
description={Solar transmissivity of the windows},
magnitude=\num{46},
unit=\unit{\percent},
}
\newcommand{\windowTransmissivity}{\gls{windowTransmissivity}\xspace}

\newglossaryentry{airCurtainLossReduction}{name=\ensuremath{\zeta_\mathrm{aircurt}},
type=constants,
sort=Zzetaaircurt,
description={Door loss reduction by air curtains},
magnitude=\num{60},
unit=\unit{\percent},
}
\newcommand{\airCurtainLossReduction}{\gls{airCurtainLossReduction}\xspace}

\newglossaryentry{interiorSolarFraction}{name=\ensuremath{\zeta_\mathrm{int}},
type=constants,
sort=Zzetacab,
description={Frac.\ of heat absorbed by interior},
magnitude=\num{30},
unit=\unit{\percent},
}
\newcommand{\interiorSolarFraction}{\gls{interiorSolarFraction}\xspace}

\newglossaryentry{roofCoverageFraction}{name=\ensuremath{\zeta_\mathrm{roof}},
type=constants,
sort=Zzetaroof,
description={Frac.\ of roof covered by components},
magnitude=\num{66},
unit=\unit{\percent},
}
\newcommand{\roofCoverageFraction}{\gls{roofCoverageFraction}\xspace}

\newglossaryentry{windowFraction}{name=\ensuremath{\zeta_\mathrm{win}},
type=constants,
sort=Zzetawin,
description={Frac.\ of walls consisting of windows},
magnitude=\num{35.4},
unit=\unit{\percent},
}
\newcommand{\windowFraction}{\gls{windowFraction}\xspace}

\newglossaryentry{gravitationalAccelaration}{name=\ensuremath{g},
type=constants,
sort=g,
description={Earth's gravitational acceleration},
magnitude=\num{9.81},
unit=\unit{\meter\per\second\squared},
}
\newcommand{\gravitationalAccelaration}{\gls{gravitationalAccelaration}\xspace}

\newglossaryentry{stefanBoltzmannConst}{name=\ensuremath{\sigma},
type=constants,
sort=Zsigma,
description={Stefan Boltzmann constant},
magnitude=\num{5.67e-08},
unit=\unit{\watt\per\meter\squared\per\kelvin\tothe{4}},
}
\newcommand{\stefanBoltzmannConst}{\gls{stefanBoltzmannConst}\xspace}

\newglossaryentry{tinit}{name=\ensuremath{t_0},
type=symbols,
sort=t0,
description={Initial time},
unit=\unit{\second},
}
\newcommand{\tinit}{\gls{tinit}\xspace}

\newglossaryentry{tfinal}{name=\ensuremath{t_\mathrm{f}},
type=symbols,
sort=tf,
description={Final time},
unit=\unit{\second},
}
\newcommand{\tfinal}{\gls{tfinal}\xspace}

\newglossaryentry{weight}{name=\ensuremath{w_i},
type=symbols,
sort=wi,
description={Weight of sample $i$.},
unit=\unit{-},
}
\newcommand{\weight}{\gls{weight}\xspace}

\newglossaryentry{PMVmax}{name=\ensuremath{\Psi_\mathrm{max}},
type=symbols,
sort=Zpsimax,
description={Maximum allowed \acs{PMV} value},
unit=\unit{-},
}
\newcommand{\PMVmax}{\gls{PMVmax}\xspace}

\newglossaryentry{PMVmin}{name=\ensuremath{\Psi_\mathrm{min}},
type=symbols,
sort=Zpsimin,
description={Minimum allowed \acs{PMV} value},
unit=\unit{-},
}
\newcommand{\PMVmin}{\gls{PMVmin}\xspace}

\newglossaryentry{FiToInt}{name=\ensuremath{F_\mathrm{\mathit{i}\rightarrow{}int}},
type=symbols,
sort=Fi2int,
description={View factor passenger surface $i$ to interior},
magnitude=\num{30},
unit=\unit{\percent},
}
\newcommand{\FiToInt}{\gls{FiToInt}\xspace}

\newglossaryentry{FiToRh}{name=\ensuremath{F_\mathrm{\mathit{i}\rightarrow{}rh}},
type=symbols,
sort=Fi2rh,
description={View factor passenger surface $i$ to \acs{RH}},
}
\newcommand{\FiToRh}{\gls{FiToRh}\xspace}

\newglossaryentry{FiToShell}{name=\ensuremath{F_\mathrm{\mathit{i}\rightarrow{}s,i}},
type=symbols,
sort=Fi2si,
description={View factor passenger surface $i$ to shell inside},
}
\newcommand{\FiToShell}{\gls{FiToShell}\xspace}

\newglossaryentry{idxi}{name=\ensuremath{i},
type=symbols,
sort=i,
description={Iteration index},
}
\newcommand{\idxi}{\gls{idxi}\xspace}

\newglossaryentry{airCurtainConsumption}{name=\ensuremath{P_\mathrm{aircurt}},
type=symbols,
sort=Paircurt,
description={Consumed air curtain power},
unit=\unit{\watt},
}
\newcommand{\airCurtainConsumption}{\gls{airCurtainConsumption}\xspace}

\newglossaryentry{Phc}{name=\ensuremath{P_\mathrm{hc}},
type=symbols,
sort=Phc,
description={Electric power consumption of the air heating and cooling unit},
}
\newcommand{\Phc}{\gls{Phc}\xspace}

\newglossaryentry{Phvac}{name=\ensuremath{P_\mathrm{hvac}},
type=symbols,
sort=Phvac1,
description={Total power consumption of the \acs{HVAC} system},
}
\newcommand{\Phvac}{\gls{Phvac}\xspace}

\newglossaryentry{PhvacTimeAverage}{name=\ensuremath{\bar{P}_\mathrm{hvac}},
type=symbols,
sort=Phvac2,
description={Cons. of the \acs{HVAC} system averaged over time},
}
\newcommand{\PhvacTimeAverage}{\gls{PhvacTimeAverage}\xspace}

\newglossaryentry{PhvacAnnualAverage}{name=\ensuremath{\breve{P}_\mathrm{hvac}},
type=symbols,
sort=Phvac3,
description={Cons. of the \acs{HVAC} system, annual average},
}
\newcommand{\PhvacAnnualAverage}{\gls{PhvacAnnualAverage}\xspace}

\newglossaryentry{PRH}{name=\ensuremath{P_\mathrm{rh}},
type=symbols,
sort=Prh,
description={Electric power consumption of the \acsp{RH}},
}
\newcommand{\PRH}{\gls{PRH}\xspace}

\newglossaryentry{Qcool}{name=\ensuremath{\dot{Q}_\mathrm{cool}},
type=symbols,
sort=Qcool,
description={Air heat removal in cooling mode},
}
\newcommand{\Qcool}{\gls{Qcool}\xspace}

\newglossaryentry{Qdoor}{name=\ensuremath{\dot{Q}_\mathrm{door}},
type=symbols,
sort=Qdoor,
description={Heat flow through open doors},
unit=\unit{\watt},
}
\newcommand{\Qdoor}{\gls{Qdoor}\xspace}

\newglossaryentry{Qhc}{name=\ensuremath{\dot{Q}_\mathrm{hc}},
type=symbols,
sort=Qhc,
description={Air heating and cooling heat flow},
}
\newcommand{\Qhc}{\gls{Qhc}\xspace}

\newglossaryentry{QhcSS}{name=\ensuremath{\dot{Q}_\mathrm{hc,ss}},
type=symbols,
sort=Qhcss,
description={Air heating and cooling heat flow in steady state},
unit=\unit{\watt},
}
\newcommand{\QhcSS}{\gls{QhcSS}\xspace}

\newglossaryentry{Qheat}{name=\ensuremath{\dot{Q}_\mathrm{heat}},
type=symbols,
sort=Qheat,
description={Air heating in heating mode},
}
\newcommand{\Qheat}{\gls{Qheat}\xspace}

\newglossaryentry{QconvInt}{name=\ensuremath{\dot{Q}_\mathrm{h,int}},
type=symbols,
sort=Qhint,
description={Convective cooling of interior},
}
\newcommand{\QconvInt}{\gls{QconvInt}\xspace}

\newglossaryentry{QconvRH}{name=\ensuremath{\dot{Q}_\mathrm{h,rh}},
type=symbols,
sort=Qhrh,
description={Convective cooling of \acsp{RH}},
}
\newcommand{\QconvRH}{\gls{QconvRH}\xspace}

\newglossaryentry{QconvShellInside}{name=\ensuremath{\dot{Q}_\mathrm{h,s,i}},
type=symbols,
sort=Qhsi,
description={Convective heating of shell inside},
}
\newcommand{\QconvShellInside}{\gls{QconvShellInside}\xspace}

\newglossaryentry{QconvShellOutside}{name=\ensuremath{\dot{Q}_\mathrm{h,s,o}},
type=symbols,
sort=Qhso,
description={Convective cooling of shell outside},
}
\newcommand{\QconvShellOutside}{\gls{QconvShellOutside}\xspace}

\newglossaryentry{Qcond}{name=\ensuremath{\dot{Q}_\mathrm{k}},
type=symbols,
sort=Qk,
description={Conductive heat loss through shell},
}
\newcommand{\Qcond}{\gls{Qcond}\xspace}

\newglossaryentry{Qother}{name=\ensuremath{\dot{Q}_\mathrm{other}},
type=symbols,
sort=Qother,
description={Other heat input to cabin},
}
\newcommand{\Qother}{\gls{Qother}\xspace}

\newglossaryentry{Qpass}{name=\ensuremath{\dot{Q}_\mathrm{pass}},
type=symbols,
sort=Qpass,
description={Heat intorudced by all passengers},
}
\newcommand{\Qpass}{\gls{Qpass}\xspace}

\newglossaryentry{QradIntToShell}{name=\ensuremath{\dot{Q}_\mathrm{r,int\rightarrow{}s,i}},
type=symbols,
sort=Qrint2si,
description={Radiative heat exchange interior to shell inside},
}
\newcommand{\QradIntToShell}{\gls{QradIntToShell}\xspace}

\newglossaryentry{QradRhToInt}{name=\ensuremath{\dot{Q}_\mathrm{r,rh\rightarrow{}int}},
type=symbols,
sort=Qrrh2int,
description={Radiative heat exchange \acs{RH} to interior},
}
\newcommand{\QradRhToInt}{\gls{QradRhToInt}\xspace}

\newglossaryentry{QradRhToShell}{name=\ensuremath{\dot{Q}_\mathrm{r,rh\rightarrow{}s,i}},
type=symbols,
sort=Qrrh2si,
description={Radiative heat exchange \acs{RH} to shell inside},
}
\newcommand{\QradRhToShell}{\gls{QradRhToShell}\xspace}

\newglossaryentry{QradShellOutside}{name=\ensuremath{\dot{Q}_\mathrm{r,s,o}},
type=symbols,
sort=Qrso,
description={Radiative heat exchange shell outside to ambient},
}
\newcommand{\QradShellOutside}{\gls{QradShellOutside}\xspace}

\newglossaryentry{clo}{name=\ensuremath{R_\mathrm{clo}},
type=symbols,
sort=Rclo,
description={Clothing insulation},
}
\newcommand{\clo}{\gls{clo}\xspace}

\newglossaryentry{Tcab}{name=\ensuremath{T_\mathrm{cab}},
type=symbols,
sort=Tcab,
description={Cabin air temperature},
}
\newcommand{\Tcab}{\gls{Tcab}\xspace}

\newglossaryentry{Tcool}{name=\ensuremath{T_\mathrm{cool}},
type=symbols,
sort=Tcool,
description={Cooling temperature setpoint},
unit=\unit{\kelvin},
}
\newcommand{\Tcool}{\gls{Tcool}\xspace}

\newglossaryentry{Theat}{name=\ensuremath{T_\mathrm{heat}},
type=symbols,
sort=Theat,
description={Heating temperature setpoint},
unit=\unit{\kelvin},
}
\newcommand{\Theat}{\gls{Theat}\xspace}

\newglossaryentry{Tint}{name=\ensuremath{T_\mathrm{int}},
type=symbols,
sort=Tint,
description={Interior temperature},
unit=\unit{\kelvin},
}
\newcommand{\Tint}{\gls{Tint}\xspace}

\newglossaryentry{TmeanRadiant}{name=\ensuremath{T_\mathrm{mr}},
type=symbols,
sort=Tmr,
description={Mean radiant temperature},
}
\newcommand{\TmeanRadiant}{\gls{TmeanRadiant}\xspace}

\newglossaryentry{TRH}{name=\ensuremath{T_\mathrm{rh}},
type=symbols,
sort=Trh,
description={\Acs{RH} temperature},
}
\newcommand{\TRH}{\gls{TRH}\xspace}

\newglossaryentry{TShellInside}{name=\ensuremath{T_\mathrm{s,i}},
type=symbols,
sort=Tsi,
description={Shell inside temperature},
}
\newcommand{\TShellInside}{\gls{TShellInside}\xspace}

\newglossaryentry{TShellOutside}{name=\ensuremath{T_\mathrm{s,o}},
type=symbols,
sort=Tso,
description={Shell outside temperature},
}
\newcommand{\TShellOutside}{\gls{TShellOutside}\xspace}

\newglossaryentry{inputVector}{name=\ensuremath{\mathbf{u}},
type=symbols,
sort=u,
description={Input vector},
}
\newcommand{\inputVector}{\gls{inputVector}\xspace}

\newglossaryentry{airCurtainUsage}{name=\ensuremath{u_\mathrm{aircurt}},
type=symbols,
sort=uaircurt,
description={Air curtain active (binary)},
unit=\unit{-},
}
\newcommand{\airCurtainUsage}{\gls{airCurtainUsage}\xspace}

\newglossaryentry{operatingMode}{name=\ensuremath{u_\mathrm{hc}},
type=symbols,
sort=uhc,
description={Operating mode of the air heating and cooling unit (integer)},
unit=\unit{-},
}
\newcommand{\operatingMode}{\gls{operatingMode}\xspace}

\newglossaryentry{rhUsage}{name=\ensuremath{u_\mathrm{rh}},
type=symbols,
sort=urh,
description={\Acs{RH} active (binary)},
unit=\unit{-},
}
\newcommand{\rhUsage}{\gls{rhUsage}\xspace}

\newglossaryentry{PMVvar}{name=\ensuremath{\Psi},
type=symbols,
sort=Zpsi,
description={\Acs{PMV}},
unit=\unit{-},
}
\newcommand{\PMVvar}{\gls{PMVvar}\xspace}

\newglossaryentry{PMVapproximation}{name=\ensuremath{\hat{\Psi}},
type=symbols,
sort=Zpsi,
description={\Acs{PMV} approximation},
unit=\unit{-},
}
\newcommand{\PMVapproximation}{\gls{PMVapproximation}\xspace}

\newglossaryentry{shadowFraction}{name=\ensuremath{\zeta_\mathrm{sh}},
type=symbols,
sort=Zzetash,
description={Shaded time fraction},
unit=\unit{-},
}
\newcommand{\shadowFraction}{\gls{shadowFraction}\xspace}

\newacronym{HEV}{HEV}{hybrid electric vehicle}
\newacronym{BEV}{BEV}{battery electric vehicle}
\newacronym{EV}{EV}{electric vehicle}
\newacronym{SOH}{SOH}{state of health}
\newacronym{SOR}{SOR}{state of resistance}
\newacronym{SOC}{SOC}{state of charge}
\newacronym{SOE}{SOE}{state of energy}
\newacronym{DOD}{DOD}{depth of discharge}
\newacronym{EOL}{EOL}{end-of-life}
\newacronym{EMS}{EMS}{energy management system}
\newacronym{PMP}{PMP}{Pontryagin's minimum principle}
\newacronym{OCP}{OCP}{optimal control problem}
\newacronym{ECMS}{ECMS}{equivalent consumption minimization strategy}
\newacronym{DP}{DP}{dynamic programming}
\newacronym{PI}{PI}{pro\-portional-inte\-gral}
\newacronym{HESS}{HESS}{hybrid energy storage system}
\newacronym{GA}{GA}{genetic algorithm}
\newacronym{ANN}{ANN}{artificial neural network}
\newacronym{PHEV}{PHEV}{plug-in hybrid electric vehicle}
\newacronym{MPC}{MPC}{model predictive control}
\newacronym{PSO}{PSO}{particle swarm optimization}
\newacronym{DC}{DC}{direct current}
\newacronym{OCV}{OCV}{open-circuit voltage}
\newacronym{VCU}{VCU}{vehicle control unit}
\newacronym{HVAC}{HVAC}{heating, ven\-ti\-la\-tion, and air-con\-di\-tio\-ning}
\newacronym{LTO}{LTO}{lithium-titanate-oxide}
\newacronym{FCEV}{FCEV}{fuel-cell electric vehicle}
\newacronym{VBZ}{VBZ}{Verkehrsbetriebe Zürich}
\newacronym{BFH}{BFH}{Bern University of Applied Sciences}
\newacronym{ODE}{ODE}{ordinary differential equation}
\newacronym{DAE}{DAE}{differential-algebraic system of equations}
\newacronym{PTC}{PTC}{positive temperature coefficient}
\newacronym{VCC}{VCC}{vapor compression cycle}
\newacronym{RH}{RH}{radiant heater}
\newacronym[
prefixfirst={a\ }, %
prefix={an\ }, %
]{NLP}{NLP}{non-linear program}
\newacronym{PMV}{PMV}{predicted mean vote}
\newacronym{PPD}{PPD}{predicted percentage dissatisfied}
\newacronym{DNI}{DNI}{direct normal irradiance}
\newacronym{DHI}{DHI}{diffuse horizontal irradiance}
\newacronym{GHI}{GHI}{global horizontal irradiance}
\newacronym{COP}{COP}{coefficient of performance}
\newacronym{AC}{AC}{air conditioning}
\newacronym[
prefixfirst={a\ }, %
prefix={an\ }, %
]{HP}{HP}{heat pump}
\newacronym{GHG}{GHG}{greenhouse gas}
\newacronym{LCA}{LCA}{life-cycle assessment}
\newacronym{WLTP}{WLTP}{world\-wide-harmo\-nized light vehicles test procedure}
\newacronym{MAE}{MAE}{mean absolute error}
\newacronym{LCC}{LCC}{life-cycle costing}
\newacronym{SFOE}{SFOE}{Swiss Federal Office of Energy}
\newacronym[
prefixfirst={a\ }, %
prefix={an\ }, %
]{LSTM}{LSTM}{long short-term memory}
\newacronym{GIS}{GIS}{geographic information system}

\newabbreviation{MeteoSwissAbbr}{Meteo\-Swiss}{Swiss Federal Office of Meteorology and Climatology}
\newcommand{\theMeteoSwiss}{\ifglsused{MeteoSwissAbbr}{}{the }\gls{MeteoSwissAbbr}\xspace} %

\begin{document}

\begin{frontmatter}

\title{Optimization of the Energy-Comfort Trade-Off of HVAC Systems in Electric City Buses Based on a Steady-State Model}

\author{Fabio Widmer}
\ead{fawidmer@idsc.mavt.ethz.ch}
\author{Stijn van Dooren}
\author{Christopher H.\ Onder}
\address{Institute for Dynamic Systems and Control, ETH Zürich, 8092 Zürich, Switzerland}

\begin{abstract}
The electrification of public transport vehicles offers the potential to relieve city centers of pollutant and noise emissions.
Furthermore, electric buses have lower life-cycle \ac{GHG} emissions than diesel buses, particularly when operated with sustainably produced electricity.
However, the \ac{HVAC} system can consume a significant amount of energy, thus limiting the achievable driving range.
In this paper, we address the \ac{HVAC} system in an electric city bus by analyzing the trade-off between the energy consumption and the thermal comfort of the passengers.
We do this by developing a dynamic thermal model for the bus, which we simplify by considering it to be in steady state.
We introduce a method that is able to quickly optimize the steady-state \ac{HVAC} system inputs for a large number of samples representative of a year-round operation.
A comparison between the results from the steady-state optimization approach and a dynamic simulation reveals small deviations in both the \ac{HVAC} system power demand and achieved thermal comfort.
Thus, the approximation of the system performance with a steady-state model is justified.
We present two case studies to demonstrate the practical relevance of the approach.
First, we show how the method can be used to compare different \ac{HVAC} system designs based on a year-round performance evaluation.
Second, we show how the method can be used to extract setpoints for online controllers that achieve close-to-optimal performance without any predictive information.
In conclusion, this study shows that a steady-state analysis of the \ac{HVAC} systems of an electric city bus is a valuable approach to evaluate and optimize its performance.
\end{abstract}

\glsresetall %


\end{frontmatter}

\section{Introduction}

\subsection{Motivation}
In the face of global warming, an increasing number of countries are committing themselves to net-zero emission targets \cite{Lang:2023ht}.
Transportation is a substantial contributor to emissions, accounting for around 20\% of global \ce{CO2} emissions, with nearly half of that originating from road-based passenger transport \cite{Ritchie:2021ov}.

Both public transport and vehicle electrification can help to decrease the carbon footprint of passenger transport.
In fact, electric buses achieve approximately 65\% less \ac{GHG} emissions than diesel buses and 75\% less than private gasoline vehicles per passenger kilometer during operation \cite{NationalAcademies:2021kg}.
Even if lifetime \ac{GHG} emissions are considered and even if a very high grid carbon intensity is assumed, electric buses are preferable over diesel-powered buses \cite{Nordelof:2019kc}.
Beyond global emissions reduction, electric propulsion in public transport can mitigate local environmental impacts in city centers, alleviating issues such as noise and pollutant emissions.
Furthermore, public transportation utilizes significantly less space per passenger kilometer compared to private vehicles \cite{Heran:2008zj}, which is an important consideration in densely populated urban areas.
For these reasons, a widespread adoption of \acp{BEV} in public transport is desirable.

While \acp{BEV} offer the aforementioned benefits, they also pose technical challenges.
One prominent issue is the high energy demand of the \ac{HVAC} system during seasons that require heating, as \acp{BEV} do not have an abundant heat source like diesel-powered vehicles.
Furthermore, public transport vehicles, in particular city buses, are typically poorly insulated and are subject to frequent openings of large doors.
For instance, the overall energy consumption of an electric city bus was shown to double during winter conditions due to the \ac{HVAC} system operation \cite{Cigarini:2021ox}.
This places heavy constraints on the achievable driving range in battery-powered applications.

Achieving a reduction in \ac{HVAC} energy consumption typically conflicts with thermal comfort requirements.
Thus, most research on \ac{HVAC} systems in public transportation vehicles focuses on enhancing the trade-off between these competing goals.
Improvements of this trade-off can be achieved along two primary paths, which can be treated independently or in conjunction: improving the \emph{design} or enhancing the \emph{control} of \ac{HVAC} systems.
In the subsequent review section, we analyze and categorize existing literature on both of these topics.

\subsection{Literature Review}

For the design and control of \ac{HVAC} systems, distinctions emerge when comparing public transportation applications to those of other domains.
\emph{Passenger vehicles} are usually operated for short periods of time.
Thus, most research focuses on transient scenarios like heat-up or cool-down \cite{Brusey:2018zj,Schaut:2020}.
This requires dynamic models, for which frameworks \cite{Docimo:2018xf} or purpose-built simulation tools \cite{Meyer:2013pl} have been suggested.
In contrast, public transportation vehicles are typically operated for extended periods of time and thus, the transient phases of heat-up and cool-down are much less relevant.
Compared to vehicular applications, thermal capacity and insulation values of \emph{buildings} are significantly higher, leading to thermal timescales on the order of hours.
Thus, models are almost uniquely dynamic \cite{bacher2011identifying,privara2013building} and predictive control is commonly used to plan the corresponding thermal trajectories in accordance with weather forecast \cite{oldewurtel2012use,drgovna2020all}.
On the other hand, transit buses exhibit a thermal timescale on the order of ten minutes \cite{Riachi:2014jr,Liebers:2017zj}.
This is significantly faster than the changing ambient conditions, such as the ambient temperature.
In summary, the long duration of driving missions and the comparably short timescales of the thermal systems in transit buses potentially allow for the application of steady-state models.

However, most research that deals with \ac{HVAC} \emph{control optimization} in public transportation applications is based on dynamic models.
For instance, dynamic models are employed to assess heuristic controller designs \cite{Amri:2011hv}, to evaluate objective functions during \ac{GA}-based optimization of control parameters \cite{AlHaddad:2022mr}, to implement multi-level \ac{MPC} approaches \cite{Hofstadter:2018mj,Dullinger:2018yhb}, or as a training basis for \ac{LSTM} neural networks that are used in a neural network \ac{MPC} algorithm \cite{Sommer:2021vu}.
In contrast, the utilization of steady-state models for controller design and optimization is rare.
Instances include the use of a static black-box model, derived from real-life measurements, to assess temperature setpoint changes \cite{Beusen:2013lg}, and the analysis of steady-state simulation results of a passenger car to generate controller setpoints \cite{Cvok:2021zt}.

In the area of \emph{design optimization}, dynamic models are also the most common option.
Several researchers propose simulation methods to estimate the energy consumption of \ac{HVAC} systems using dynamic models \cite{Riachi:2014jr,Dullinger:2015ub,Vepsalainen:2019sn,Basma:2020hn,Wang:2020je}.
While these publications do not directly show comparisons between system designs, the models are typically motivated by such objectives.
Some researchers directly compare the performance of different \ac{HVAC} systems based on dynamic models in buses \cite{Gohlich:2015,Pathak:2020th} or trains \cite{Barone:2020xu}.
With a few exceptions, steady-state models are rarely used.
For instance, models from first principles are evaluated in steady-state to suggest an extension of the \ac{WLTP} to include \ac{HVAC} loads \cite{Mansour:2018jz} or to evaluate the effectiveness of system components like \acp{HP} and \acp{RH} \cite{Widmer:2023df}.
Alternatively, based on measurement data, data-driven steady-state models can be developed for the \ac{HVAC} system consumption within a larger energy consumption estimation framework \cite{Szilassy:2022vg}, or even for the entire vehicle consumption calculation, which enables evaluating various design changes \cite{Wang:2021hy}.

Due to their inherently high computational demands, dynamic models, whether used for control or design purposes, are inadequate to analyze year-round operation.
For this reason, usually only simplified problems involving a single scenario are considered \cite{Dullinger:2018yhb,Cvok:2021zt,Pathak:2020th}.
Many researchers attempt to approximate the year-round operation by considering a small sample of up to ten scenarios, e.g., \cite{Schaut:2020,Amri:2011hv,AlHaddad:2022mr,Hofstadter:2018mj,Dullinger:2015ub,Basma:2020hn,Wang:2020je}.
These scenarios are typically based on heuristic approaches or expert knowledge, e.g., by selecting representative days for each season \cite{Amri:2011hv} or for nine different weather classes \cite{Dullinger:2015ub} or \enquote{typical} days for each month \cite{Wang:2020je}.
An alternate approach is to base the scenario selection on machine learning tools.
For instance, some researchers suggest using clustering algorithms to find representative scenarios \cite{Luger:2016ht,Li:2020hy}.
Only a limited number of researchers use dynamic models to fully consider year-round influences.
This can be achieved by simulating every day over an entire year \cite{Barone:2020xu} or by using a reference year and categorizing days based on their average temperatures to limit the number of simulations \cite{Gohlich:2015}.
An alternate approach is to avoid frequent evaluations of a computationally expensive model by creating a surrogate energy consumption model whose parameters are tuned with Monte Carlo simulations \cite{Vepsalainen:2019sn}.
Similarly, a surrogate model-based optimization of the charging infrastructure under consideration of all loads, including \ac{HVAC} system consumption, is suggested in \cite{Foda:2023cu}, allowing the optimization to be based on a large number of driving missions.

Conversely, the comparatively low computational demand of steady-state models facilitates the evaluation of the annual performance.
To this end, various methods can be used, such as Monte Carlo simulations \cite{Beusen:2013lg}, weighted averages with the frequency of different temperature ranges \cite{Mansour:2018jz}, or computing hourly averages for realistic operating hours \cite{Widmer:2023df,Wang:2021hy}.

\subsection{Research Statement}

In previous research, some researchers have suggested the application of steady-state models for the \ac{HVAC} system of public transportation vehicles.
Utilizing a steady-state thermal model rather than a dynamic one significantly reduces the computational time required for simulations or for solving optimization problems.
This efficiency enhancement facilitates the evaluation of year-round performance.
However, to the best of our knowledge, none of the published research based on steady-state models provides quantitative comparisons or evidence why such an approach is appropriate.
In this paper, we fill this gap by conducting a quantitative comparison between steady-state optimization and dynamic simulation results.
Furthermore, we explore the possibility of leveraging steady-state models for both system design comparisons and control improvements.

\subsection{Paper Structure}

This paper is structured as follows.
In \cref{sec:system-modeling}, we introduce the thermal model of a city bus cabin and its \ac{HVAC} system.
We explain how we quantify thermal comfort, introduce the disturbances and their data sources, and introduce the steady-state approximation of the model.
In \cref{sec:optimization}, we formulate the optimization problem based on the steady-state model and present a method to solve it.
In \cref{sec:results}, we present and discuss the results.
Finally, in \cref{sec:conclusion}, we draw conclusions and make suggestions for further research.

\section{System Modeling}
\label{sec:system-modeling}

In this section, we first introduce a dynamic thermal model of the passenger cabin and the \ac{HVAC} system of an electric city bus.
We then explain how we quantify thermal comfort and introduce the data sources used for our case studies.
Finally, we introduce a steady-state approximation of the model, which forms the basis for optimizing the \ac{HVAC} operation as described in \cref{sec:optimization}.
All model constants are listed in \cref{sec:model-constants}.

\subsection{Dynamic Thermal Model}
\label{sec:dynamic-thermal-model}

\begin{figure}
\centering
\small
\def\svgwidth{\linewidth}
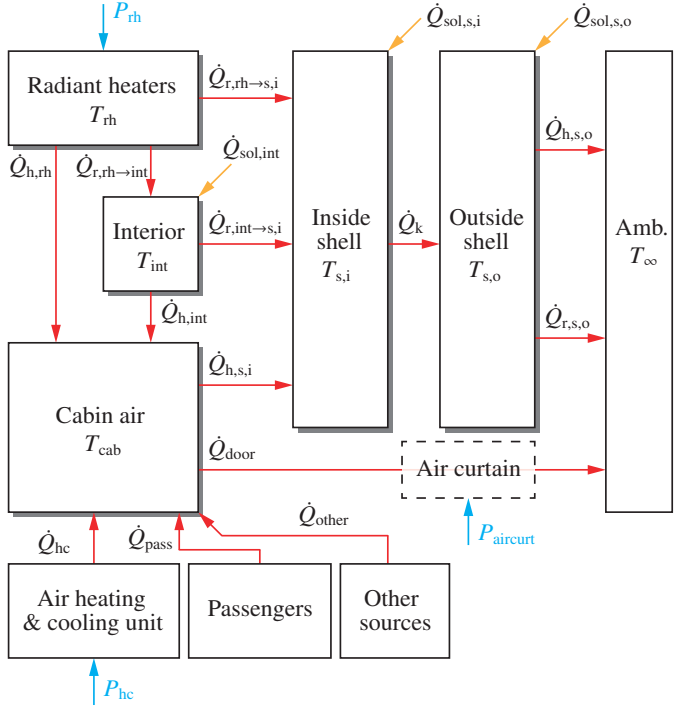
\caption{
Model component overview.
Thermal reservoirs are represented as shaded blocks.
Red arrows denote heat flows, yellow arrows denote solar heat gain, and blue arrows denote electric power flows.
Time dependencies are omitted for clarity.
}
\label{fig:heat-flow-overview}
\end{figure}

\Cref{fig:heat-flow-overview} provides an overview of the thermal model of the passenger cabin and the \ac{HVAC} system of an electric city bus.
It depicts all model components along with all the heat and power flows.

The bus cabin and its shell are modeled using five thermal reservoirs, which are characterized by their respective temperature.
The cabin air temperature $\Tcab(t)$ represents the thermal environment for the passengers.
For the sake of simplicity, we model the entire cabin with one uniform temperature and do not separately treat the driver compartment.
The bus interior, which includes the seats and hand rails, is modeled with one lumped thermal capacity with temperature $\Tint(t)$.
As the interior is in convective heat exchange with the surrounding cabin air, it has a smoothing effect on the cabin air temperature during transients, reducing the impact of sudden changes in heat flows, such as those caused by door openings.
We model the bus shell with two separate components, namely an \emph{inside} shell and an \emph{outside} shell with temperatures $\TShellInside(t)$ and $\TShellOutside(t)$, respectively.
The reason for this distinction is that the inside shell surface temperature is relevant for the radiation heat exchange with the passengers, as explained in the comfort model below.
Finally, the \acp{RH} are modeled as surfaces with a uniform temperature $\TRH(t)$.

The \ac{HVAC} system that is used to provide thermal comfort for the passengers consists of an air heating and cooling unit and \acp{RH}, which are electrically heated surfaces at a temperature \TRH.
The \acp{RH} increase the thermal comfort by directly heating passengers via radiation, without significantly increasing the air temperature \cite{Widmer:2023df}.
Lastly, air curtains are also considered, which can reduce the losses through open doors.
The inputs to the \ac{HVAC} system can be summarized as follows:
\begin{equation}
\inputVector(t) = \big(\Phc(t),\, \PRH(t),\, \operatingMode(t),\, \airCurtainUsage(t),\, \rhUsage(t) \big), \\
\end{equation}
where $\Phc(t)$ and $\PRH(t)$ denote the power provided to the air heating and cooling unit and the \acp{RH}, respectively.
The integer variable $\operatingMode(t)$ is used to define the operating mode of the air heating and cooling unit:
\begin{equation}
\operatingMode(t) = \begin{cases}
 1 & \text{heating,} \\
 0 & \text{passive,} \\
-1 & \text{cooling.}
\end{cases}
\label{eq:operating-modes}
\end{equation}
The binary variables $\airCurtainUsage(t)$ and $\rhUsage(t)$ are used to enable or disable the air curtains and the \acp{RH}, respectively.
The total power consumption of the \ac{HVAC} system is given by
\begin{equation}
\Phvac(t) = \Phc(t) + \PRH(t) + \airCurtainConsumption(t) \,,
\label{eq:phvactot}
\end{equation}
where $\airCurtainConsumption(t)$ denotes the power consumption of the air curtains.

The thermal model is subject to various disturbances, which are summarized in the vector
\begin{equation}
\disturbanceVector(t) = \left(\Tamb(t),\, \beamIrradiance(t),\, \diffuseIrradiance(t),\, \timeVar,\, \Npass(t),\, \doorOpenFraction(t) \right) .
\end{equation}
The uniform ambient temperature $\Tamb(t)$, the \ac{DNI} $\beamIrradiance(t)$, and the \ac{DHI} $\diffuseIrradiance(t)$ characterize the meteorological conditions, which influence the various heat gains and losses shown in \cref{fig:heat-flow-overview}.
The time variable \timeVar denotes an absolute point in time.
It is explicitly included in the disturbance vector, since it is used to calculate solar angles and shadow factors further below.
The number of passengers $\Npass(t)$ and the fraction of open doors $\doorOpenFraction(t)$ influence metabolic heating rate and the door losses, respectively.

In the following subsections, we first introduce the models for all the heat flows and then state the system of \acp{ODE} that describe the temperature dynamics.

\subsubsection{Convection, Conduction, Radiation}

The modeled temperature differences lead to convective, conductive, and radiative heat exchange.
We apply Newton's law of cooling to obtain the convective heat transfer between the shell and cabin air, the interior and cabin air, the shell and ambient, and the \acp{RH} and cabin air, respectively:
\begin{align}
\QconvShellInside(t) &= \hInside \cdot \Ahull \cdot (\Tcab(t) - \TShellInside(t)) , \\
\QconvInt(t) &= \hInside \cdot \Aint \cdot (\Tint(t) - \Tcab(t)) , \\
\QconvShellOutside(t) &= \hOutside \cdot \Ahull \cdot (\TShellOutside(t) - \Tamb(t)) ,  \\
\QconvRH(t) &= \hRh \cdot \ARH \cdot (\TRH(t) - \Tcab(t)) .
\end{align}
The values for the convective heat transfer coefficients at the \acp{RH} \hRh, inside \hInside, and outside \hOutside are based on the norm \cite{ISO6946:2017bv} and are adjusted for increased air movement inside the cabin and an average bus velocity of \qty{15}{\kilo \meter \per \hour}.
The surface area of the hull \Ahull is given by the cabin dimensions (\cabinLength, \cabinWidth, \cabinHeight).
The \ac{RH} surface area \ARH is given by the single panel size and by assuming a distribution of 16 panels on the ceiling of the cabin \cite{Widmer:2023df}.
We tune the interior surface area \Aint to reproduce heat-up and cool-down trajectories generated with a three-dimensional simulation model implemented in a high-fidelity simulator.

For the conductive heat transfer through the shell, we again use Newton's law of cooling, i.e.,
\begin{equation}
\Qcond(t) = \conductance \cdot \Ahull \cdot (\TShellInside(t) - \TShellOutside(t)) ,
\end{equation}
where \conductance is the mean heat conductance through the bus enclosure.
This value is calculated to match an overall U-value of \qty{2.9}{\watt \per \meter \squared \per \kelvin}, which has been determined for a city bus in a climate chamber \cite{Sidler:2019hg}.

For simplicity, all surfaces are assumed to be perfect emitters and the windows are assumed to be opaque to infrared radiation \cite{Incropera:2007as}.
Assuming a uniform ambient temperature, the heat emitted by the shell of the bus is given by
\begin{equation}
\QradShellOutside(t) = \stefanBoltzmannConst \cdot \Ahull \cdot \left(\TShellOutside(t)^4 - \Tamb(t)^4\right) ,
\end{equation}
where \stefanBoltzmannConst is the Ste\-fan-Boltz\-mann constant.
In the bus cabin, radiative exchange occurs between the \acp{RH}, the shell, and the interior.
To characterize this exchange, we need the view factors between these surfaces \cite{Incropera:2007as}.
Since the \acp{RH} are mounted in a coplanar arrangement on the ceiling, their mutual view factor is zero.
Furthermore, without specifying a specific geometry, we assume that the view factor from the interior to itself is negligible.
For the view factor from the \acp{RH} to the interior, we assume a value of $\FrhToInt = \glsmagnitude{FrhToInt}\glsunit{FrhToInt}$.
By applying the \enquote{summation} and \enquote{reciprocity} rules \cite{Incropera:2007as}, we can then determine both the values of the view factors from the \acp{RH} to the shell \FrhToShell and from the interior to the shell \FIntToShell.
Finally, the radiative heat exchange between the \acp{RH} and shell, \acp{RH} and interior, and interior and shell, respectively, is calculated as:
\begin{align}
\QradRhToShell(t) &= \stefanBoltzmannConst \cdot \ARH \cdot \FrhToShell \cdot (\TRH(t)^4 - \TShellInside(t)^4) , \\
\QradRhToInt(t) &= \stefanBoltzmannConst \cdot \ARH \cdot \FrhToInt \cdot (\TRH(t)^4 - \Tint(t)^4) , \\
\QradIntToShell(t) &= \stefanBoltzmannConst \cdot \Aint \cdot \FIntToShell \cdot (\Tint(t)^4 - \TShellInside(t)^4) .
\end{align}

\subsubsection{Door Losses}

To estimate the buoyancy-induced air exchange through opened doors, we use a model based on Bernoulli's law and the ideal gas law \cite{Schalin:1998ds}.
Based on the air exchange, we can calculate the corresponding heat losses through the doors with height \doorHeight and a combined overall width \totalDoorWidth as follows:
\begin{multline}
\Qdoor(t) = \frac{\airDensity \cdot \dischargeCoefficient \cdot \sqrt{\gravitationalAccelaration \cdot \doorHeight^3} \cdot  \totalDoorWidth}{3} \cdot \sqrt{\frac{| \Tcab(t) - \Tamb(t) |}{\Tamb(t)}} \\
\cdot (\Tcab(t) - \Tamb(t)) \cdot \airHeatCapacity \cdot \doorOpenFraction(t) \cdot (1 - \airCurtainLossReduction \cdot \airCurtainUsage(t)) ,
\label{eq:qdoor}
\end{multline}
where 
\airDensity denotes the air density,
\gravitationalAccelaration denotes the gravitational acceleration, and
\dischargeCoefficient denotes an empirical discharge coefficient \cite{Schalin:1998ds}.
The fraction $\doorOpenFraction(t)$ denotes the fraction of the doors that are open.
For instance, $\doorOpenFraction(t) = 1$ means all doors are open.
The final term concerns the air curtains, which are activated with the binary variable $\airCurtainUsage(t)$, and reduce the losses by a constant fraction \airCurtainLossReduction, the value of which is based on \cite{Pathak:2020th,Schalin:1998ds}.

The air curtain power requirement is given by 
\begin{equation}
\airCurtainConsumption(t) = \airCurtainUsage(t) \cdot \doorOpenFraction(t) \cdot \airCurtainSpecification ,
\end{equation}
where the power consumption of the blowers \airCurtainSpecification is based on measurements on a single door in an experimental study that features a comparable vehicle \cite{Sidler:2021vu}.
The power requirement of the air curtains is zero if they are disabled (i.e., $\airCurtainUsage(t)=0$) or if the doors are closed (i.e., $\doorOpenFraction(t)=0$).

\subsubsection{Solar Heat Gain}

For the calculation of the solar heat gain, we consider both direct and diffuse irradiation.
The solar altitude angle $\solarAltitude(t)$ can be calculated based on the local coordinates and the datetime \cite{solar-position-calculator}.
Based on the \ac{DNI} denoted by $\beamIrradiance(t)$ and \ac{DHI} denoted by $\diffuseIrradiance(t)$, the mean irradiance on the roof and on the wall can be calculated as
\begin{align}
\roofIrradiance(t) &= \cos \left(\frac{\pi}{2} - \solarAltitude(t)\right) \cdot \beamIrradiance(t) + \diffuseIrradiance(t) , \\
\meanWallIrradiance(t) &= \frac{\cos \left(\solarAltitude(t)\right)}{\pi} \cdot \beamIrradiance(t) + \frac{1}{2} \cdot \diffuseIrradiance(t) ,
\end{align}
where the latter is calculated based on the assumption that the bus is driving in all directions with equal probability  \cite{Widmer:2023df}.

\begin{figure}
\centering
\begin{tikzpicture}

\begin{axis}[%
width=0.98\linewidth,
height=0.5\linewidth,
xmin=0,
xmax=24,
ymin=0,
ymax=1,
xlabel={Time of day},
x unit={\unit{\hour}},
ylabel={\shadowFraction},
y unit={-},
xticklabel={\pgfkeys{/pgf/number format/hour code={\tick}}},
xtick={0,6,12,18,24},
legend style={at={(0.5, 1)}, anchor=south, legend columns=3, draw=none, fill=none,font=\small,},
]

\addplot [color1, thick, mark=*, mark size=1pt]
  table[]{img/shadow-fraction/shadow-fraction-3-x10-Dec-2022.tsv};
\addlegendentry{December 10}

\addplot [color3, thick, mark=*, mark size=1pt]
  table[]{img/shadow-fraction/shadow-fraction-2-x21-Mar-2022.tsv};
\addlegendentry{March 21}

\addplot [color2, thick, mark=*, mark size=1pt]
  table[]{img/shadow-fraction/shadow-fraction-1-x24-Jul-2022.tsv};
\addlegendentry{July 24}

\end{axis}
\end{tikzpicture}%
\caption{
Shadow fraction \shadowFraction averaged over a specific bus route in Zürich, based on a simulation using \ac{GIS} software and a three-dimensional model of the city of Zürich \cite{Zurich3d,swissALTI3D}.
}
\label{fig:shadow-fraction}
\end{figure}
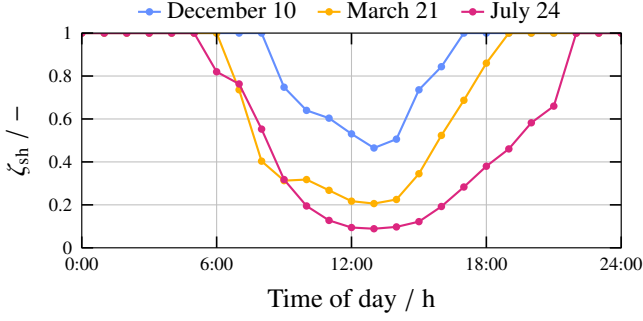

Based on the irradiance values, the total heat gain on the outside shell of the cabin is given by
\begin{multline}
\QsolarShellOutside(t) = (1 - \shadowFraction(t)) \cdot \Big( \roofArea \cdot \roofIrradiance(t) \cdot \surfaceAbsorptivity \cdot (1 - \roofCoverageFraction) \\
+ \wallArea \cdot \meanWallIrradiance(t) \cdot (1 - \windowFraction) \cdot \surfaceAbsorptivity \Big) ,
\end{multline}
where \roofArea and \wallArea denote the roof surface area and the combined surface area of all walls, respectively.
The surfaces of the bus hull are painted in bright colors and their solar absorptivity is denoted by \surfaceAbsorptivity, which is estimated based on data published in \cite{Incropera:2007as}.
The fractions \roofCoverageFraction and \windowFraction denote the fraction of the roof that is shaded by components such as batteries, and the fraction of the walls that are windows, respectively.
For the fraction $\shadowFraction(t)$, which represents the fraction of time in which the bus is in the shade, we have developed a simulation in \ac{GIS} software to calculate the portion of the road along a route in Zürich that is shaded according to a three-dimensional model of the city, including buildings and trees \cite{Zurich3d}, and topography \cite{swissALTI3D}.
The model is then evaluated for every full hour over an entire year.
We use linear interpolation between the calculated values to determine the value of the shadow fraction $\shadowFraction(t)$ at any specific point in time.
Some sample results of this process are shown in \cref{fig:shadow-fraction}.

A part of the radiation hitting the windows is assumed to be transmitted into the cabin, characterized by the solar transmissivity \windowTransmissivity of the windows, whose value is given by the manufacturer.
The solar radiation transmitted into the bus through the windows is assumed to be fully absorbed by either the interior or the inside shell.
The corresponding heat flows are:
\begin{align}
\QsolarInt(t) &= (1 - \shadowFraction(t)) \cdot \wallArea \cdot \meanWallIrradiance(t) \cdot \windowFraction \cdot \windowTransmissivity \cdot \interiorSolarFraction , \\
\QsolarShellInside(t) &= (1 - \shadowFraction(t)) \cdot \wallArea \cdot \meanWallIrradiance(t) \cdot \windowFraction \cdot \windowTransmissivity \cdot (1 - \interiorSolarFraction) ,
\end{align}
where we assume that $\interiorSolarFraction = \glsmagnitude{interiorSolarFraction}\glsunit{interiorSolarFraction}$ is absorbed by the interior.

\subsubsection{Heat Sources}

The power provided to the \acp{RH} is denoted by $\PRH(t)$, which we assume to be fully converted into heat.
Furthermore, we assume that the \acp{RH} are insulated against the ceiling.

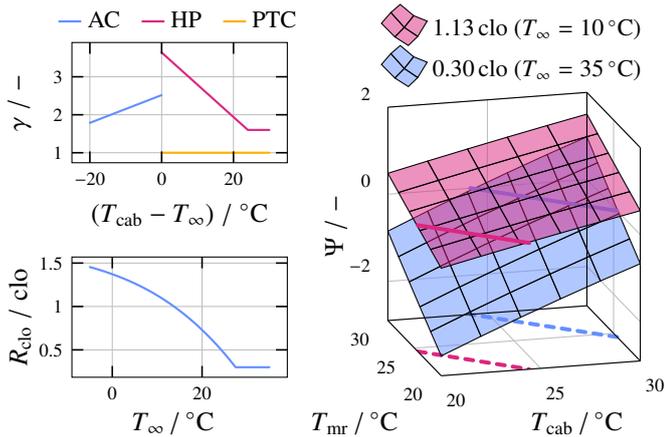
\begin{figure}
\centering
\begin{tikzpicture}

\begin{axis}[%
width=0.32\linewidth,
height=0.18\linewidth,
at={(0,0)},
name=ax1,
scale only axis,
xlabel={$(\Tcab - \Tamb)$},
x unit = {\unit{\celsius}},
ylabel={\COPvar},
y unit = -,
legend style={at={(0.5, 1)}, anchor=south, legend columns=3, draw=none, fill=none,font=\small,},
y label style={ at={(-0.4cm,0.5)} },
]
\addplot [color=color1, thick]
  table[]{img/model-parameters/model-parameters-1-cop-ac.tsv};
\addlegendentry{\acs{AC}}

\addplot [color=color2, thick]
  table[]{img/model-parameters/model-parameters-2-cop-hp.tsv};
\addlegendentry{\acs{HP}}

\addplot [color=color3, thick]
  table[]{img/model-parameters/model-parameters-3-cop-ptc.tsv};
\addlegendentry{\acs{PTC}}

\end{axis}

\begin{axis}[%
width=0.32\linewidth,
height=0.18\linewidth,
at={($(0, -0.18\linewidth) + (0, -1.25cm)$)},
scale only axis,
ylabel={\clo},
y unit = clo,
y label style={ at={(-0.4cm,0.5)} },
]
\addplot [color=color1, thick]
  table[]{img/model-parameters/model-parameters-4-clo.tsv};
\end{axis}

\pgfplotsset{ytick style={draw=none}}
\pgfplotsset{xtick style={draw=none}}
\pgfplotsset{ztick style={draw=none}}

\begin{axis}[%
width=0.375\linewidth,
height=0.15\linewidth + 0.15\linewidth + 1.1cm,
at={($(0.3\linewidth, -0.18\linewidth) + (1.5cm, -1.23cm)$)},
scale only axis,
tick align=outside,
zlabel={\PMVvar},
z unit = -,
zlabel style={
    at={(-0.4cm,0.5)}
},
view={-15}{15},
legend style={at={(0.5, 1)}, anchor=south, legend columns=1, draw=none, fill=none,font=\small,},
zmajorgrids,
reverse legend,
zmin = -2.9,
zmax = 2,
]

\addplot3 [color=color1, dashed, line width=1.5pt, forget plot]
 table[] {img/model-parameters/model-parameters-7.tsv};
\addplot3 [color=color2, dashed, line width=1.5pt, forget plot]
 table[] {img/model-parameters/model-parameters-10.tsv};

\addplot3[%
surf,
fill opacity=0.5, fill=color1, faceted color=black, z buffer=sort, mesh/rows=6]
table[point meta=\thisrow{c}] {%
img/model-parameters/model-parameters-5-x0-30clo-30-00C.tsv};
\addlegendentry{\qty{0.30}{clo} ($\Tamb=\qty{35}{\celsius}$)}

\addplot3 [color=color1, line width=1.5pt, forget plot]
 table[] {img/model-parameters/model-parameters-6.tsv};
 
\addplot3[%
surf,
fill opacity=0.5, fill=color2, faceted color=black, z buffer=sort, mesh/rows=6]
table[point meta=\thisrow{c}] {%
img/model-parameters/model-parameters-8-x1-13clo-10-00C.tsv};
\addlegendentry{\qty{1.13}{clo} ($\Tamb=\qty{10}{\celsius}$)}

\addplot3 [color=color2, line width=1.5pt, forget plot]
 table[] {img/model-parameters/model-parameters-9.tsv};
 \end{axis}

\node[at={(0.15\linewidth, -3.45cm)}] {\Tamb{} / \unit{\celsius}};
\node[at={(0.42\linewidth, -3.45cm)}] {\TmeanRadiant{} / \unit{\celsius}};
\node[at={(0.75\linewidth, -3.45cm)}] {\Tcab{} / \unit{\celsius}};
\end{tikzpicture}%
\caption{
Visualization of the temperature-dependent model components.
The solid lines in the right plot correspond to the temperature values representing a thermally neutral environment ($\PMVvar(t) = 0$).
The dashed lines are the corresponding projections onto the ground plane, for clarity.
}
\label{fig:model-parameters}
\end{figure}

In steady state, the heat provided by the air heating and cooling unit can be calculated based on the unit's \ac{COP} $\COPvar(t)$:
\begin{equation}
\QhcSS(t) = \operatingMode(t) \cdot \COPvar(t) \cdot \Phc(t) ,
\label{eq:steady-state-hc}
\end{equation}
where the mode selection $\operatingMode(t)$ determines the sign of the heat flow, as introduced in \cref{eq:operating-modes}.
For purely resistive heating (e.g., with \ac{PTC} elements), the \ac{COP} is $\COPvar(t) = 1$ in heating mode.
For heating based on \pgls{HP} and for cooling, the modeled \ac{COP} depends on the temperature difference between the two thermal reservoirs.
The corresponding dependency, as shown in the top left plot of \cref{fig:model-parameters}, is fitted to measurement data of various operating points provided by the \ac{VCC} manufacturer.
This fit achieves an $R^2$ value of more than 90\%.

Due to the thermal dynamics of the \ac{VCC}, as well as the dynamics of the low-level controller of the expansion valve, the heat flow of \pgls{HP} or an \ac{AC} system does not change instantaneously when the compressor power $\Phc(t)$ is changed.
For simplicity, we model these dynamics by filtering the steady-state values $\QhcSS(t)$ as given in \cref{eq:steady-state-hc} with a first-order element, similarly to, e.g., \cite{Merino:2012nt}.
This results in an \ac{ODE} for the heat provided by the air heating and cooling unit:
\begin{equation}
\timeDerivative \Qhc(t) = \frac{1}{\timeConstVcc} \cdot \left(\QhcSS(t) - \Qhc(t)\right) .
\label{eq:vcc-ode}
\end{equation}
Based on a physical simulation of a \ac{VCC}, we have determined a time constant $\timeConstVcc = \glsmagnitude{timeConstVcc}\,\glsunit{timeConstVcc}$ to be a reasonable value, which is slightly longer than what is observed in units for passenger cars \cite{Salazar:2014zs,Rostiti:2015mz}.

By assuming that the heat supplied by the air heating and cooling unit is fully transferred into the cabin air, we implicitly assume that the unit is operated in recirculation mode, which is a reasonable assumption for city buses, where the air exchange through open doors typically provides enough fresh air to maintain sufficient cabin air quality.

The heat emitted by $\Npass(t)$ passengers is modeled as
\begin{equation}
\Qpass(t) = \Npass(t) \cdot \metabolicHeat ,
\label{eq:metabolic-heat}
\end{equation}
where \metabolicHeat denotes the average metabolic heat rate of a human when seated \cite{ISO7730:2005mj}.
Lastly, all other heat sources (electric components, lights, screens) are captured by a constant heat flow \Qother.

\subsubsection{Thermal Reservoirs}

Based on the first law of thermodynamics and the heat flows shown in \cref{fig:heat-flow-overview}, we can formulate the following system of \acp{ODE} for the five thermal reservoirs:
\begin{subequations}\label{eq:odes}
\begin{align}
\timeDerivative \TRH &= \frac{1}{\heatCapacityRH} \left( \PRH - \QconvRH - \QradRhToInt - \QradRhToShell \right) \label{eq:rh-balance}, \\
\timeDerivative \Tint &= \frac{1}{\heatCapacityInterior} \left( \QsolarInt + \QradRhToInt - \QconvInt - \QradIntToShell \right) \label{eq:int-balance}, \\
\timeDerivative \Tcab &= \frac{1}{\heatCapacityCabin} \Big( \Qpass + \Qother + \Qhc + \QconvRH + \QconvInt \notag \\
& \hspace{3cm} - \QconvShellInside - \Qdoor \Big) \label{eq:cab-balance} , \\
\timeDerivative \TShellInside &= \frac{1}{\heatCapacityShellInside} \left( \QconvShellInside + \QsolarShellInside + \QradRhToShell + \QradIntToShell - \Qcond \right) \label{eq:si-balance}, \\
\timeDerivative \TShellOutside &= \frac{1}{\heatCapacityShellOutside} \left( \Qcond + \QsolarShellOutside - \QconvShellOutside - \QradShellOutside \right) , \label{eq:so-balance}
\end{align}
\end{subequations}
where we omit the time dependencies of the temperature and heat flow variables to increase readability.

We estimate the heat capacity of the \acp{RH} \heatCapacityRH based on dimensions and manufacturer's specifications.
The heat capacity of the cabin air \heatCapacityCabin is calculated based on the cabin volume \cabinVolume, as
\begin{equation}
\heatCapacityCabin = \airDensity \cdot \airHeatCapacity \cdot \cabinVolume .
\end{equation}
The values of the heat capacity for the interior \heatCapacityInterior, the shell inside \heatCapacityShellInside, and outside \heatCapacityShellOutside are determined to reproduce heat-up and cool-down trajectories as mentioned above.

\subsection{Comfort}

We model thermal comfort according to the norm \cite{ISO7730:2005mj}, which characterizes the thermal comfort in terms of the \ac{PMV}, ranging from $-3$ (\enquote{cold}) to $+3$ (\enquote{hot}).
This model has been originally proposed by Fanger \cite{Fanger:1970jh} and is the most widely used for bus cabins \cite{dasNeves:2020je}.
For our calculations, we assume 
a constant air velocity $\vCab = \glsmagnitude{vCab}\,\glsunit{vCab}$ and 
a constant relative humidity of $\rhCab = \glsmagnitude{rhCab}\glsunit{rhCab}$.
For the metabolic heat rate, we use $\metabolicHeat = \glsmagnitude{metabolicHeat}\,\glsunit{metabolicHeat}$, which corresponds to a value of \qty{1.2}{met}.
Thus, the \ac{PMV} is calculated according to \cite{ISO7730:2005mj} as a function of the air temperature $\Tcab(t)$, the mean radiant temperature $\TmeanRadiant(t)$, and the clothing insulation factor $\clo(\Tamb(t))$, i.e.:
\begin{equation}
\PMVvar(t) = f( \Tcab(t), \TmeanRadiant(t), \clo(\Tamb(t)), \vCab, \rhCab, \metabolicHeat) .
\label{eq:pmv}
\end{equation}
Two sample visualizations of this dependency are shown in the right plot of \cref{fig:model-parameters}.
The insulation factor $\clo(\Tamb(t))$ quantifies the clothing of the passengers, which depends on the ambient temperature \cite{Havenith:2012dh}.
We use the same dependency as in a previous study \cite{Widmer:2023df}, shown in the bottom left plot of \cref{fig:model-parameters}.

To determine the mean radiant temperature $\TmeanRadiant(t)$, we describe a passenger by an upright cuboid with the dimensions \qtyproduct{0.25 x 0.25 x 1.7}{\meter}, which does not take part in the radiation balance of the enclosure.
The passenger can thus be considered a \enquote{passive} sensory device, apart from its metabolic heat release described by \cref{eq:metabolic-heat}.
As passenger placement is typically not recorded, we use a randomized approach for passenger placement in our model.
For a visualization of such a distribution, we refer to our previous publication \cite{Widmer:2023df}.
Again, using the view factor relations given in \cite{Incropera:2007as}, the mean radiant temperature of a passenger is denoted by
\begin{multline}
\TmeanRadiant(t)^4 = \frac{1}{{\sum_{\idxi} \Ai} } \cdot 
\Bigg( \TRH(t)^4 \cdot \sum_{\idxi} (\Ai \cdot \FiToRh) \\
+
\TShellInside(t)^4 \cdot \sum_{\idxi} (\Ai \cdot \FiToShell)
+ 
\Tint(t)^4 \cdot \sum_{\idxi} (\Ai \cdot \FiToInt) \Bigg)
\end{multline}
where the index \idxi is used to iterate over the five surfaces of the cuboid.
The view factors to the \acp{RH}, \FiToRh, can be determined analytically, as explained in \cite{Gross:1981jy}.
For the view factors to the interior, we assume $\FiToInt = \glsmagnitude{FiToInt}\glsunit{FiToInt}$.
The view factor to the shell, \FiToShell, can then be found using the summation rule \cite{Incropera:2007as}.
Finally, to obtain an overall estimate of the \ac{PMV} for all passengers, we average the respective \ac{PMV} values as calculated in \cref{eq:pmv}.

\subsection{Data Sources}
\label{sec:data-sources}

The numerical data needed to characterize the disturbances $\disturbanceVector(t)$ is collected from multiple sources.
The operating schedule, the passenger load, and the door openings are obtained from the ZTBus dataset \cite{Widmer:2023zj}, where we use the data for bus number 183.
Concerning door openings, the ZTBus dataset only contains information of whether at least one door is open.
Our analysis of log data from a similar vehicle reveals that, on average, about 60\% of the four doors open at each stop.
We include this factor of 60\% into the pre-processing of $\doorOpenFraction(t)$ for all the simulations.
All meteorological data are obtained from a meteorological station of \theMeteoSwiss located in Zürich.

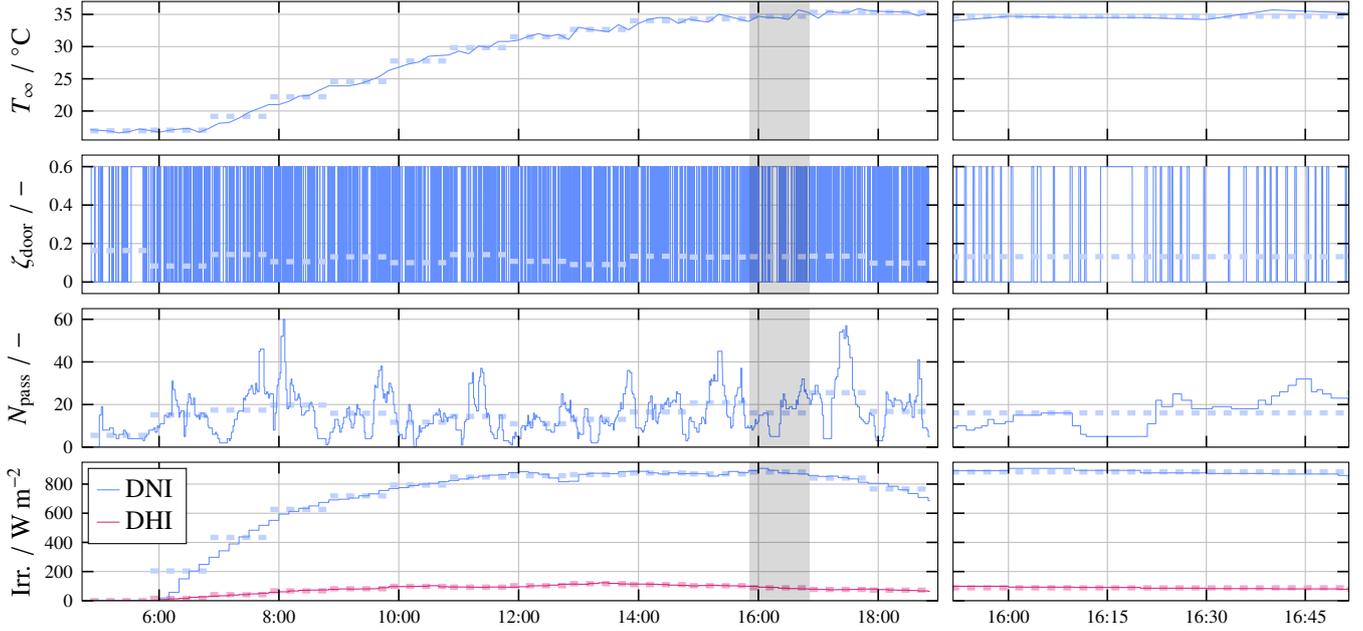
\begin{figure*}
\centering
\begin{tikzpicture}

\pgfplotsset{
    leftAxesStyle/.style={
		width=0.7\linewidth,
	},
	rightAxesStyle/.style={
		width=0.37\linewidth,
		xmin=15.87,
		xmax=16.85,
		xtick={16,16.25,16.5,16.75},
	},
	temperatureAxLims/.style={
		ymin=15.5,
		ymax=37,
	},
	nPassAxLims/.style={
		ymin=0,
		ymax=65,
	},
	irradAxLims/.style={
		ymin=0,
		ymax=950,
	},
}

\newcommand{\shadeBlack}{%
\fill[black, opacity=0.15] (axis cs:15.86, \pgfkeysvalueof{/pgfplots/ymin}) rectangle (axis cs:16.86, \pgfkeysvalueof{/pgfplots/ymax});
}

\begin{groupplot}[
	group style={
		group size=2 by 4,
		vertical sep=0.2cm,
		horizontal sep=0.2cm,
		xlabels at=edge bottom,
		xticklabels at=edge bottom,
		ylabels at=edge left,
		yticklabels at=edge left,
	},
    height=0.1863\linewidth,
    xticklabel={\pgfkeys{/pgf/number format/hour code={\tick}}},
	y label style={ at={(-0.5cm,0.5)} },
	enlarge x limits = 0.01,
]

\nextgroupplot[
leftAxesStyle,
temperatureAxLims,
ylabel={\Tamb},
y unit = {\unit{\celsius}},
]

\addplot [color1!40!white, staticStyle]
  table[]{img/dynamic-comparison-hot/dynamic-comparison-hot-2-T--amb-Static.tsv};

\addplot [color1, dynamicStyle]
  table[]{img/dynamic-comparison-hot/dynamic-comparison-hot-1-T--amb-Dyn.tsv};

\shadeBlack

\nextgroupplot[
rightAxesStyle,
temperatureAxLims,
]

\addplot [color1!40!white, staticStyle]
  table[]{img/dynamic-comparison-hot/dynamic-comparison-hot-2-T--amb-Static.tsv};

\addplot [color1, dynamicStyle]
  table[]{img/dynamic-comparison-hot/dynamic-comparison-hot-1-T--amb-Dyn.tsv};

\nextgroupplot[
leftAxesStyle,
ylabel={\doorOpenFraction},
y unit = {-},
]
\addplot [color1, dynamicStyle]
  table[]{img/dynamic-comparison-hot/dynamic-comparison-hot-3-doorDyn.tsv};

\addplot [color1!40!white, staticStyle]
  table[]{img/dynamic-comparison-hot/dynamic-comparison-hot-4-doorStatic.tsv};

\shadeBlack

\nextgroupplot[rightAxesStyle]

\addplot [color1, dynamicStyle]
  table[]{img/dynamic-comparison-hot/dynamic-comparison-hot-3-doorDyn.tsv};

\addplot [color1!40!white, staticStyle]
  table[]{img/dynamic-comparison-hot/dynamic-comparison-hot-4-doorStatic.tsv};

\nextgroupplot[
leftAxesStyle,
nPassAxLims,
ylabel={\Npass},
y unit = {-},
]

\addplot [color1!40!white, staticStyle]
  table[]{img/dynamic-comparison-hot/dynamic-comparison-hot-6-N-pStatic.tsv};

\addplot [color1, dynamicStyle]
  table[]{img/dynamic-comparison-hot/dynamic-comparison-hot-5-N-pDyn.tsv};

\shadeBlack

\nextgroupplot[
rightAxesStyle,
nPassAxLims,
]

\addplot [color1!40!white, staticStyle]
  table[]{img/dynamic-comparison-hot/dynamic-comparison-hot-6-N-pStatic.tsv};

\addplot [color1, dynamicStyle]
  table[]{img/dynamic-comparison-hot/dynamic-comparison-hot-5-N-pDyn.tsv};

\nextgroupplot[
leftAxesStyle,
irradAxLims,
ylabel = {Irr.},
y unit = {\unit{\watt\per\square\meter}},
legend style={\legendNorthWest}
]

\addplot [color1!40!white, staticStyle]
  table[]{img/dynamic-comparison-hot/dynamic-comparison-hot-8-DNIStatic.tsv};

\addplot [color2!40!white, staticStyle]
  table[]{img/dynamic-comparison-hot/dynamic-comparison-hot-10-DHIStatic.tsv};

\addplot [color1, dynamicStyle]
  table[]{img/dynamic-comparison-hot/dynamic-comparison-hot-7-DNIDyn.tsv};
\addlegendentry{\acs{DNI}}

\addplot [color2]
  table[]{img/dynamic-comparison-hot/dynamic-comparison-hot-9-DHIDyn.tsv};
\addlegendentry{\acs{DHI}}
\shadeBlack

\nextgroupplot[
rightAxesStyle,
irradAxLims,
xticklabels={16:00,16:15,16:30,16:45},
]

\addplot [color1!40!white, staticStyle]
  table[]{img/dynamic-comparison-hot/dynamic-comparison-hot-8-DNIStatic.tsv};

\addplot [color2!40!white, staticStyle]
  table[]{img/dynamic-comparison-hot/dynamic-comparison-hot-10-DHIStatic.tsv};

\addplot [color1, dynamicStyle]
  table[]{img/dynamic-comparison-hot/dynamic-comparison-hot-7-DNIDyn.tsv};

\addplot [color2]
  table[]{img/dynamic-comparison-hot/dynamic-comparison-hot-9-DHIDyn.tsv};

\end{groupplot}
\end{tikzpicture}%
\caption{
Disturbances $\disturbanceVector(t)$ on a summer day, recorded on \DTMdisplaydate{2019}{7}{24}{-1}.
Thin solid lines represent the time-resolved data, while thick dashed lines represent hourly averages.
The shaded segment is the basis for the visualization in \cref{fig:single-scenario-hot}.
The plots shown on the right are a zoomed version of this segment.
}
\label{fig:data-sample-hot}
\end{figure*}

\Cref{fig:data-sample-hot} shows an example of the disturbances, recorded on a summer day.
An example of a winter day is shown in \cref{fig:dynamic-comparison-cold} in the results section.

\subsection{Steady-State Approximation}
\label{sec:steady-state-approximation}

The dynamic thermal model introduced in the previous subsections can be simplified based on the principle of timescale separation.
For our system, this principle means that the dynamics of the \acp{ODE} \cref{eq:vcc-ode,eq:odes} can be neglected (i.e., all time derivatives are set to zero) if the typical rate of change of the disturbances $\disturbanceVector(t)$ is significantly lower than the characteristic response time of the system\footnote{For a first-order, linear time-invariant system, this would be the time constant.}.
In that case, the disturbances can be assumed constant over a certain period of time, and the resulting system of nonlinear equations can be evaluated using a suitable numerical method.
To accommodate for changes in the disturbances, this evaluation can be repeated with a representative sample of (constant) disturbances.
For instance, instead of simulating the entire time-resolved trajectory of around \qty{14}{\hour} shown in \cref{fig:data-sample-hot}, the nonlinear system of equations can be evaluated 14 times, if hourly averages for each of the disturbances are used as representative samples.
The resulting performance metrics, such as the \ac{HVAC} system power consumption, can then be averaged to obtain an approximation of the corresponding dynamic performance.
Such an approximation requires much less computation than the corresponding dynamic simulation.

Looking at the disturbances shown in \cref{fig:data-sample-hot}, we observe that the ambient temperature and both types of solar irradiance change only slowly and can thus be assumed constant during one hour.
Changes in the number of passengers and door openings, on the other hand, happen on a timescale very similar to or faster than the characteristic response time of the thermal dynamics, which is in the order of around \qty{10}{\minute} \cite{Riachi:2014jr,Liebers:2017zj}.
In other words, there is no clear timescale separation for these disturbances and a steady-state approximation cannot represent the dynamic operation of the \ac{HVAC} subject to these disturbances.
However, as the cabin temperature is typically closed-loop controlled, we can assume that these disturbances are rejected by the corresponding controller.
Hence, we use the steady-state approximation to evaluate the \emph{average} operation of the \emph{closed-loop controlled} \ac{HVAC} system.
The accuracy of this approximation is evaluated below in \cref{sec:ResultsValSS}.

\section{Optimal \Glsfmtshort{HVAC} Operation}
\label{sec:optimization}

Optimizing the \ac{HVAC} operation based on a dynamic thermal model is typically a challenging and computationally intensive task.
It involves the formulation of an \ac{OCP}, which can be transcribed to \pgls{NLP}, which is then solved using a suitable method.
As the sampling time must be small enough to capture the dynamics, but the driving missions are typically very long, the number of optimization variables in such \pgls{NLP} is large.
Therefore, the computation time to solve such problems is high, especially if they involve integer decision variables such as $\operatingMode(t)$, $\airCurtainUsage(t)$, and $\rhUsage(t)$.
This prohibits the evaluation of year-round performance within a reasonable time frame.
Conversely, optimizing the \ac{HVAC} operation based on the steady-state approximation described in \cref{sec:steady-state-approximation} can be accomplished by formulating a comparably small \ac{NLP}.
As this involves much less optimization variables, the solution can be computed much quicker, which enables the evaluation of year-round performance.
In this section, we formulate such an optimization problem and explain how it can be solved efficiently.

\subsection{Problem Formulation}

The objective is to minimize the total power consumption \Phvac given by \cref{eq:phvactot} for a given set of averaged disturbances \disturbanceVector while respecting a specified comfort requirement.
This can be mathematically formulated as follows:
\begin{subequations}\label{eq:optproblem}%
\begin{alignat}{2}
\underset{\inputVector}{\text{minimize}} &\quad&& \Phvac , \label{eq:min-obj} \\
\text{s.t.} &&& \text{\cref{eq:vcc-ode,eq:odes} in steady state,} \label{eq:min-balance}\\
&&& \PMVvar \in \left[ \PMVmin, \PMVmax \right] , \label{eq:pmv-req}\\
&&& 0 = \rhUsage \cdot (\TRH - \TRHTarget) , \label{eq:rh-target}\\
&&& 0 = (1 - \rhUsage) \cdot \PRH , \label{eq:rh-req}
\end{alignat}%
\end{subequations}%
where \cref{eq:pmv-req} defines the comfort requirement in the form of a box constraint on the \ac{PMV}.
Depending on the value of the binary variable \rhUsage, one of the two vanishing constraints, \cref{eq:rh-target} and \cref{eq:rh-req}, is trivially satisfied, while the other constraint is \enquote{activated}:
If the \acp{RH} are used, i.e., $\rhUsage = 1$, the \ac{RH} temperature is constrained by \cref{eq:rh-target} according to the data sheet.
If they are disabled, i.e., $\rhUsage = 0$, the corresponding power must be zero as enforced by \cref{eq:rh-req}.
This formulation prevents solutions where the \acp{RH} are operated outside of their specifications.
We omit all time dependencies in \cref{eq:optproblem} to highlight the fact the the system is assumed to be in steady state.

\subsection{Solution Approach}

The optimization problem \cref{eq:optproblem} presents three challenges for numerical solution algorithms.
In this section, we show how we address these challenges.

First, we address the three integer optimization variables \operatingMode, \rhUsage, and \airCurtainUsage.
To optimize the operating mode, we introduce two lifting variables to represent the bidirectional heat provided by the air heating and cooling unit as
\begin{equation}
\Qhc = \Qheat + \Qcool ,
\label{eq:qhvac-relaxation}
\end{equation}
where $\Qheat > 0$ denotes the heat provided by the heating system and $\Qcool < 0$ denotes the heat removal by the \ac{AC} system.
In steady state, the \ac{ODE} \cref{eq:vcc-ode} trivially resolves to $\Qhc = \QhcSS$.
Hence, using the two lifting variables and \cref{eq:steady-state-hc}, the air heating and cooling power conversion can be reformulated as follows:
\begin{equation}
\Phc = \frac{\Qheat}{\COPHP} - \frac{\Qcool}{\COPAC} \,,
\label{eq:relaxed-phvac}
\end{equation}
where \COPHP and \COPAC denote the \ac{COP} for heating and cooling, respectively (see \cref{fig:model-parameters}).
Although this reformulation allows simultaneous heating and cooling, the optimization algorithm will avoid such suboptimal behavior.
The two binary decision variables \airCurtainUsage and \rhUsage are eliminated by solving the continuous optimization problem for each of the four combinations of these variables and selecting the solution with the lowest objective value.

\begin{figure}
\centering
\begin{tikzpicture}

\pgfplotsset{colormap={myColormap}{color=(color1), color=(white), color=(color2)}}

\begin{groupplot}[
	group style={
		group size=2 by 1,
		horizontal sep=1.5cm,
	},
	width=0.5\linewidth, 
	height=0.5\linewidth,
]

\nextgroupplot[
xlabel={$(\Tcab - \Tamb)$},
x unit= K,
ylabel={$\sqrt{| \Tcab - \Tamb |} $},
y unit= K,
xmin = -5,
xmax = 5,
ymin = 0,
legend style={at={(0.5, 1)}, anchor=south, legend columns=2, draw=none, fill=none, font=\small,},
]
\addplot [color=color1, thick,
on layer=axis foreground,
]
  table[]{img/model-approximations/model-approximations-1-exact.tsv};
\addlegendentry{exact}
\addplot [color=color2, thick,
on layer=axis foreground,
]
  table[]{img/model-approximations/model-approximations-2-approx.tsv};
\addlegendentry{approx.\ \cref{eq:sqrt-of-abs-approx}} %

\nextgroupplot[
xlabel={\Tcab},
x unit = \unit{\celsius},
ylabel={\TmeanRadiant},
y unit = \unit{\celsius},
colormap name = {myColormap},
point meta min=-0.02,
point meta max=0.02,
colorbar sampled,
colorbar horizontal,
colorbar style={
	at={(0, 1.3)},
	anchor=south west,
	height=0.3cm,
	xlabel={$(\PMVvar - \PMVapproximation)$},
	xlabel style={at={(0.5,1.1)},anchor=south},
	x unit = {-},
	scaled ticks=false, %
	xticklabel style={/pgf/number format/fixed,}, %
},
view={0}{90},
ytick={10,20,30},
grid=none,
]

\addplot3[
contour filled={number=20,},
patch type=bilinear,
mesh/cols=27,
mesh/ordering=y varies,
]
table {img/model-approximations/contour-data.dat};

\end{groupplot}

\useasboundingbox(0, 0.5\linewidth);

\end{tikzpicture}%
\caption{Visualizations of the model approximations.
The left plot shows the approximation \cref{eq:sqrt-of-abs-approx}.
The right contour plot shows the error of the approximation \cref{eq:pmv-approx} for an ambient temperature of $\Tamb = \qty{10}{\celsius}$.
}
\label{fig:model-approximations}
\end{figure}
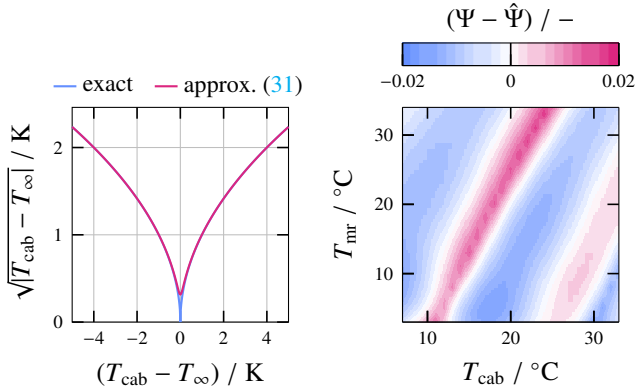

Second, the model of the door losses given by \cref{eq:qdoor} is non-smooth in the variable \Tcab.
Such non-smoothness often leads to numerical issues when using derivative-based optimization methods.
For this reason, we replace the square root in \cref{eq:qdoor} by a smooth approximation:
\begin{equation}
\sqrt{| \Tcab - \Tamb |} \approx \sqrt[4]{(\Tcab - \Tamb)^2 + 0.01} .
\label{eq:sqrt-of-abs-approx}
\end{equation}
The value $0.01$ results in good convergence behavior of the numerical method employed while keeping the error of the approximation low, as shown in the left plot of \cref{fig:model-approximations}.

Third, the relationship \cref{eq:pmv} is based on iterative calculations \cite{ISO7730:2005mj,Fanger:1970jh}.
As derivative-based optimization tools require closed-form expressions of all constraints, we use a feedforward \ac{ANN} with a single hidden layer with five neurons and the hyperbolic tangent activation function to approximate this dependency:
\begin{equation}
\PMVapproximation = \pmvAnnFunction (\Tcab,\, \TmeanRadiant) .
\label{eq:pmv-approx}
\end{equation}
The parameters of $\pmvAnnFunction(\cdot)$ are fitted for each value of $\clo (\Tamb)$, which is indicated by the subscript \Tamb.
The error of one of these approximations is visualized in the right plot of \cref{fig:model-approximations}.
The \ac{MAE} in this example is less than 0.01.

Based on these simplifications and reformulations, the optimization problem \cref{eq:optproblem} can be formulated as a set of four \acp{NLP}.
To solve these problems, we use the CasADi \cite{Andersson:2018zj} interface to Ipopt \cite{Waechter:2006:xo} and the linear solver MUMPS \cite{Amestoy:2011ze}.
On a standard laptop, each of the four \acp{NLP} can be solved in around \qty{40}{\milli\second} on average. %
In a previous publication \cite{Widmer:2023df}, we have presented an alternative solution approach which is based on a root-finding algorithm.

\section{Results}
\label{sec:results}

The result section is split into three subsections.
First, we compare the results of the optimization based on the steady-state model to simulations of the full dynamic model.
Second, we present two case studies to show how the approach can be useful for different purposes.
In the first case study, we compare the year-round performance of different \ac{HVAC} systems.
In the second case study, we extract setpoints from the optimal solutions of the steady-state models and use these setpoints for a causal controller in a dynamic simulation environment.

\subsection{Validation of the Steady-State Assumption}
\label{sec:ResultsValSS}

In this subsection, we compare the results obtained by the approach based on optimizations of the steady-state model outlined in \cref{sec:optimization} to results obtained by performing a simulation of the dynamic thermal model introduced in \cref{sec:dynamic-thermal-model}.
Thereby, we show that the steady-state model can be used to estimate the average performance of the closed-loop controlled \ac{HVAC} system of a city bus.
The following steps outline how the results shown in the subsequent figures are obtained:
\begin{enumerate}
\item We split the driving mission into segments of one hour length.
We calculate the average values for the disturbances $\disturbanceVector(t)$ in each of these segments to obtain a set of samples each representing one hour of the original mission.

\item We optimize the \ac{HVAC} operation in steady-state based on the approach presented in \cref{sec:optimization} for each of the samples.

\item We perform a dynamic simulation for the same mission.
We use the results of the steady-state optimizations within the corresponding \qty{1}{\hour}-segments as follows:
The integer variables \airCurtainUsage, \rhUsage, \operatingMode are directly used as constant inputs in a feed-forward manner in the corresponding \qty{1}{\hour}-segment.
The cabin temperature is used as a setpoint trajectory for a \ac{PI} feedback controller which regulates $\Phc(t)$.
Its proportional gain is \qty{2000}{\watt\per\kelvin} and its integral time constant is \qty{100}{\second}.
For the initial conditions of all the temperatures, we use the bus depot temperature, which we assume to depend linearly on the ambient temperature:
\begin{equation}
\depotTemperature(t) = (\Tamb(t) - \qty{20}{\celsius}) \cdot 0.5 + \qty{20}{\celsius} .
\label{eq:depot-temperature}
\end{equation}
\end{enumerate}

\begin{figure*}
\centering
\begin{tikzpicture}

\newcommand{\shadeBlack}{%
\fill[black, opacity=0.15] (axis cs:17.58, \pgfkeysvalueof{/pgfplots/ymin}) rectangle (axis cs:18.58, \pgfkeysvalueof{/pgfplots/ymax});
}

\begin{groupplot}[
	group style={
		group size=1 by 7,
		vertical sep=0.2cm,
		xlabels at=edge bottom,
		xticklabels at=edge bottom,
	},
	width=\linewidth, 
    height=0.1863\linewidth,
    xticklabel={\pgfkeys{/pgf/number format/hour code={\tick}}},
    y label style={ at={(-0.5cm,0.5)} },
	enlarge x limits = 0.01,
]

\nextgroupplot[
ylabel={\Tamb},
y unit = {\unit{\celsius}},
]

\addplot [color1!40!white, staticStyle]
  table[]{img/dynamic-comparison-cold/dynamic-comparison-cold-2-T--amb-Static.tsv};

\addplot [color1, dynamicStyle]
  table[]{img/dynamic-comparison-cold/dynamic-comparison-cold-1-T--amb-Dyn.tsv};

\shadeBlack

\nextgroupplot[
ylabel={\doorOpenFraction},
y unit = {-},
]
\addplot [color1, dynamicStyle]
  table[]{img/dynamic-comparison-cold/dynamic-comparison-cold-3-doorDyn.tsv};

\addplot [color1!40!white, staticStyle]
  table[]{img/dynamic-comparison-cold/dynamic-comparison-cold-4-doorStatic.tsv};

\shadeBlack

\nextgroupplot[
ylabel={\Npass},
y unit = {-},
]

\addplot [color1!40!white, staticStyle]
  table[]{img/dynamic-comparison-cold/dynamic-comparison-cold-6-N-pStatic.tsv};

\addplot [color1, dynamicStyle]
  table[]{img/dynamic-comparison-cold/dynamic-comparison-cold-5-N-pDyn.tsv};

\shadeBlack

\nextgroupplot[
y unit = {\unit{\watt\per\square\meter}},
ylabel = {Irr.},
legend style={\legendNorthEast}
]

\addplot [color1!40!white, staticStyle]
  table[]{img/dynamic-comparison-cold/dynamic-comparison-cold-8-DNIStatic.tsv};

\addplot [color2!40!white, staticStyle]
  table[]{img/dynamic-comparison-cold/dynamic-comparison-cold-10-DHIStatic.tsv};

\addplot [color1, dynamicStyle]
  table[]{img/dynamic-comparison-cold/dynamic-comparison-cold-7-DNIDyn.tsv};
\addlegendentry{\acs{DNI}}

\addplot [color2, dynamicStyle]
  table[]{img/dynamic-comparison-cold/dynamic-comparison-cold-9-DHIDyn.tsv};
\addlegendentry{\acs{DHI}}

\shadeBlack

\nextgroupplot[
y unit = {\unit{\celsius}},
ylabel = {Temp.},
legend style={\legendNorthEast}
]

\addplot [color1, dynamicStyle]
  table[]{img/dynamic-comparison-cold/dynamic-comparison-cold-11-cabinDyn.tsv};
\addlegendentry{\Tcab}

\addplot [color2, dynamicStyle]
  table[]{img/dynamic-comparison-cold/dynamic-comparison-cold-13-shellInsideDyn.tsv};
\addlegendentry{\TShellInside}

\addplot [color1!40!white, staticStyle]
  table[]{img/dynamic-comparison-cold/dynamic-comparison-cold-12-cabinStatic.tsv};

\addplot [color2!40!white, staticStyle,
forget plot]
  table[]{img/dynamic-comparison-cold/dynamic-comparison-cold-14-shellInsideStatic.tsv};

\shadeBlack

\nextgroupplot[
ylabel={\PMVvar},
y unit = {-},
]

\addplot [color1, dynamicStyle]
  table[]{img/dynamic-comparison-cold/dynamic-comparison-cold-17-PMVDyn.tsv};

\addplot [color1!40!white, staticStyle]
  table[]{img/dynamic-comparison-cold/dynamic-comparison-cold-18-PMVStatic.tsv};

\shadeBlack

\nextgroupplot[
ylabel={\Phvac},
y unit = {kW},
]

\addplot [color1, dynamicStyle]
  table[]{img/dynamic-comparison-cold/dynamic-comparison-cold-19-P--hvac-Dyn.tsv};

\addplot [color1!40!white, staticStyle]
  table[]{img/dynamic-comparison-cold/dynamic-comparison-cold-20-P--hvac-Static.tsv};

\shadeBlack

\end{groupplot}
\end{tikzpicture}%
\caption{
Full-day simulation of a winter day recorded on \DTMdisplaydate{2022}{12}{10}{-1}.
Thin solid lines represent dynamic simulation, while thick dashed lines represent the results of the steady-state approach proposed in this paper.
The shaded segment is the basis for the visualization in \cref{fig:single-scenario-cold}.
}
\label{fig:dynamic-comparison-cold}
\end{figure*}
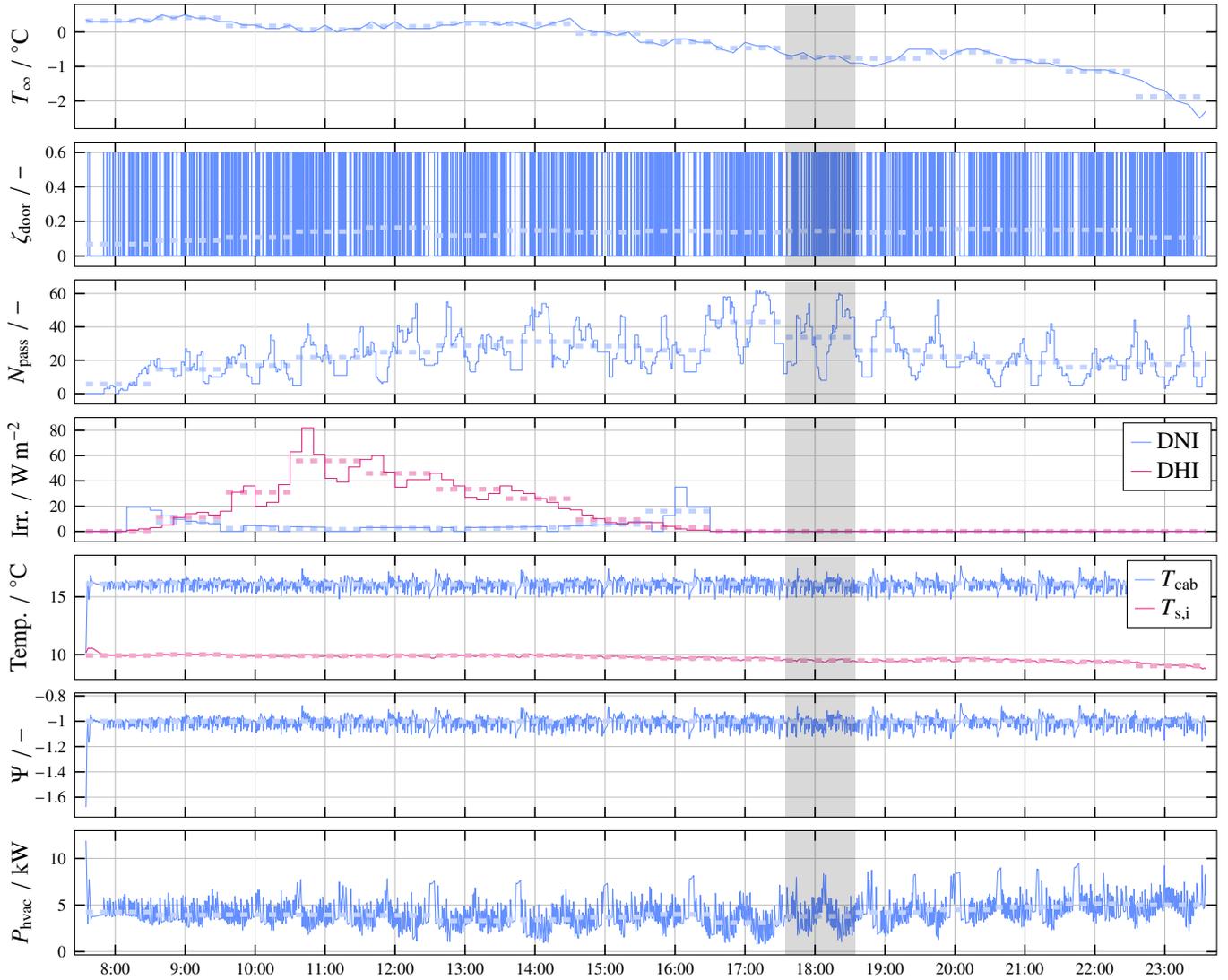
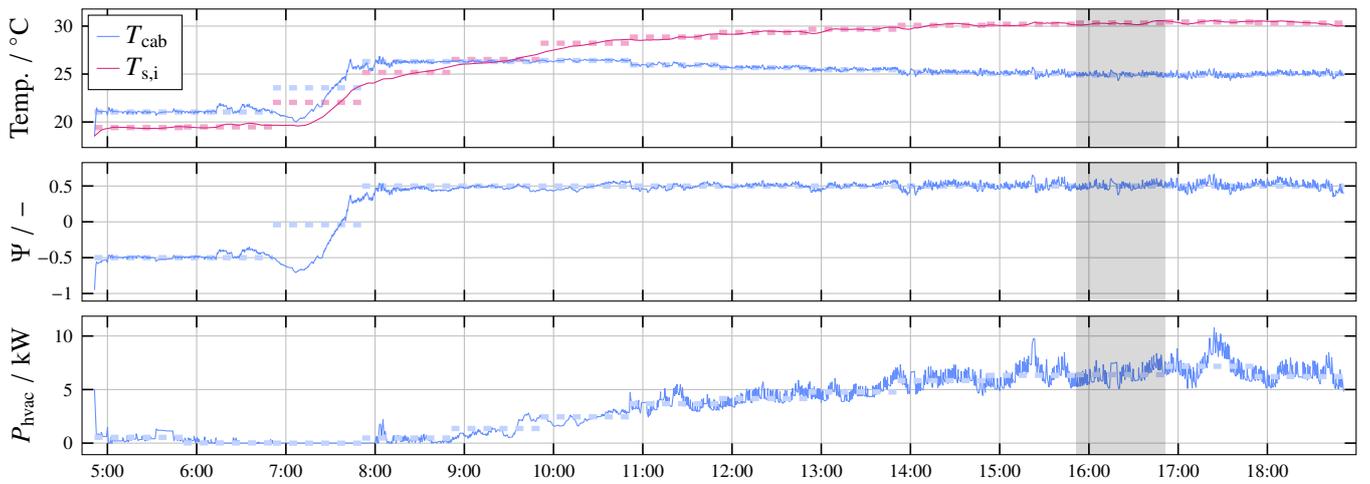
\begin{figure*}
\centering
\begin{tikzpicture}

\newcommand{\shadeBlack}{%
\fill[black, opacity=0.15] (axis cs:15.86, \pgfkeysvalueof{/pgfplots/ymin}) rectangle (axis cs:16.86, \pgfkeysvalueof{/pgfplots/ymax});
}

\begin{groupplot}[
	group style={
		group size=1 by 3,
		vertical sep=0.2cm,
		xlabels at=edge bottom,
		xticklabels at=edge bottom,
	},
	width=\linewidth, 
    height=0.1863\linewidth,
    xticklabel={\pgfkeys{/pgf/number format/hour code={\tick}}},
	y label style={ at={(-0.5cm,0.5)} },
	enlarge x limits = 0.01,
]

\nextgroupplot[
y unit = {\unit{\celsius}},
ylabel = {Temp.},
legend style={\legendNorthWest}
]

\addplot [color1!40!white, staticStyle]
  table[]{img/dynamic-comparison-hot/dynamic-comparison-hot-12-cabinStatic.tsv};

\addplot [color2!40!white, staticStyle]
  table[]{img/dynamic-comparison-hot/dynamic-comparison-hot-14-shellInsideStatic.tsv};

\addplot [color1, dynamicStyle]
  table[]{img/dynamic-comparison-hot/dynamic-comparison-hot-11-cabinDyn.tsv};
\addlegendentry{\Tcab}

\addplot [color2, dynamicStyle]
  table[]{img/dynamic-comparison-hot/dynamic-comparison-hot-13-shellInsideDyn.tsv};
\addlegendentry{\TShellInside}

\shadeBlack

\nextgroupplot[
ylabel={\PMVvar},
y unit = {-},
]

\addplot [color1!40!white, staticStyle]
  table[]{img/dynamic-comparison-hot/dynamic-comparison-hot-18-PMVStatic.tsv};

\addplot [color1, dynamicStyle]
  table[]{img/dynamic-comparison-hot/dynamic-comparison-hot-17-PMVDyn.tsv};

\shadeBlack

\nextgroupplot[
ylabel={\Phvac},
y unit = {kW},
]

\addplot [color1, dynamicStyle]
  table[]{img/dynamic-comparison-hot/dynamic-comparison-hot-19-P--hvac-Dyn.tsv};

\addplot [color1!40!white, staticStyle]
  table[]{img/dynamic-comparison-hot/dynamic-comparison-hot-20-P--hvac-Static.tsv};

\shadeBlack

\end{groupplot}
\end{tikzpicture}%
\caption{
Full-day simulation of a summer day recorded on \DTMdisplaydate{2019}{7}{24}{-1}, the disturbance trajectories of which are shown in \cref{fig:data-sample-hot}.
Thin solid lines represent dynamic simulation, while thick dashed lines represent the results of the steady-state approach proposed in this paper.
The shaded segment is the basis for the visualization in \cref{fig:single-scenario-hot}.
}
\label{fig:dynamic-comparison-hot}
\end{figure*}

\Cref{fig:dynamic-comparison-cold} shows the resulting trajectories of a full-day winter mission.
The top four plots visualize the disturbances $\disturbanceVector(t)$.
As the figure shows, the \ac{PI} controller is able to track the cabin temperature $\Tcab(t)$ well.
The frequent door openings lead to significant disturbances in the overall heat balance, which cannot be fully compensated by the controller and are thus visible in short-term deviations from the target temperature.
Even though only the cabin temperature is controlled, the shell inside temperature, which is also relevant for the thermal comfort, is shown to almost perfectly match the hourly values predicted by the steady-state solution.
As the temperatures closely follow the steady-state solution, so does the \ac{PMV}.
While door openings lead to violations from the comfort requirement of \PMVRangeTime{1}, these violations are only short in duration and small in magnitude.
The trajectory of $\Phvac(t)$ shows how slow disturbances, like changing solar irradiation or ambient temperature, lead to slow changes in the required power throughout the day, which are also present in the steady-state optimization.
As the cabin temperature is controlled in closed loop, faster disturbances like varying passenger numbers lead to temporary deviations in $\Phvac(t)$ from the corresponding steady-state estimation.

\Cref{fig:dynamic-comparison-hot} shows a comparison for a full-day summer mission.
The more demanding \ac{PMV} requirement of \PMVRangeTime{0.5} for this example is also satisfied reasonably well.
The largest violation of the target \ac{PMV} is observed after the mode switch from the \enquote{cooling} to the \enquote{passive} mode at around \DTMdisplaytime{6}{50}{0}.
Just after the switch to the \enquote{passive} mode, the \ac{PMV} starts to drop below the lower limit of $-0.5$.
Hence, to satisfy the \ac{PMV} requirements, the \enquote{heating} should have been active for a few more minutes.
From this observation, we conclude that the ideal time instance for mode switches cannot be predicted accurately with the steady-state approach, as such a switch can only happen once per hour with this approach.
In \cref{sec:case-study-2-causal-online-control}, we propose an improved method to handle mode switches in an online application.
Door air curtains are only used after about \DTMdisplaytime{10}{50}{0}, which is evident from the ripple effect appearing in $\Phvac(t)$ from that time onward.

\begin{figure}
\centering
\begin{tikzpicture}

\begin{axis}[%
width=\linewidth,
height=0.7\linewidth,
xlabel={$\max \left| \PMVvar \right|$},
x unit = {-},
ylabel={\PhvacTimeAverage},
y unit = {kW},
legend style={\legendNorthEast},
grid=none,
]
\addplot [coolingColor!40!white, 
dashed, 
mark=x, 
mark options={solid}, 
thick]
  table[]{img/dynamic-comparison-pareto/dynamic-comparison-pareto-1-ColdMission-Static.tsv};
\addlegendentry{steady-state winter}

\addplot [coolingColor, 
mark=o, 
thick,
]
  table[]{img/dynamic-comparison-pareto/dynamic-comparison-pareto-2-ColdMission-Dynamic.tsv};
\addlegendentry{dynamic winter}

\addplot [heatingColor!40!white, 
dashed, 
mark=x, 
mark options={solid}, 
thick]
  table[]{img/dynamic-comparison-pareto/dynamic-comparison-pareto-3-HotMission-Static.tsv};
\addlegendentry{steady-state summer}

\addplot [heatingColor, 
mark=o, 
mark options={solid}, 
thick]
  table[]{img/dynamic-comparison-pareto/dynamic-comparison-pareto-4-HotMission-Dynamic.tsv};
\addlegendentry{dynamic summer}

\draw[pmvWindowMarkingRectangleStyle] (axis cs:-0.06,3.87) rectangle (axis cs:0.4,9.08);
\node[pmvWindowMarkingLabelStyle,below=-1pt, anchor=north] at (axis cs:0.1,3.87) {$\PMVvar=0$};

\draw[pmvWindowMarkingRectangleStyle] (axis cs:0.44,3.2) rectangle (axis cs:0.72,6.38);
\node[pmvWindowMarkingLabelStyle,below=-1pt, anchor=north] at (axis cs:0.55,3.2) {$\PMVRange{0.5}$};

\draw[pmvWindowMarkingRectangleStyle] (axis cs:0.94,2.63) rectangle (axis cs:1.12,4.35);
\node[pmvWindowMarkingLabelStyle,above=-3pt, anchor=south] at (axis cs:1.25,4.4) {$\PMVRange{1}$};

\draw[pmvWindowMarkingRectangleStyle] (axis cs:1.44,2.17) rectangle (axis cs:1.62,2.81);
\node[pmvWindowMarkingLabelStyle,below=-1pt, anchor=north] at (axis cs:1.3,2.2) {$\PMVRange{1.5}$};

\draw[pmvWindowMarkingRectangleStyle] (axis cs:1.94,0.98) rectangle (axis cs:2.12,2.37);
\node[pmvWindowMarkingLabelStyle,above=-3pt, anchor=south] at (axis cs:2.0,2.52) {$\PMVRange{2}$};

\end{axis}
\end{tikzpicture}%
\caption{
Comparison of Pareto fronts obtained using the steady-state approach and the dynamic simulation, for the summer and winter missions shown in \cref{fig:dynamic-comparison-cold,fig:dynamic-comparison-hot}.
The gray dotted boxes and corresponding labels denote the comfort range used to optimize the steady-state model.
}
\label{fig:dynamic-comparison-pareto}
\end{figure}
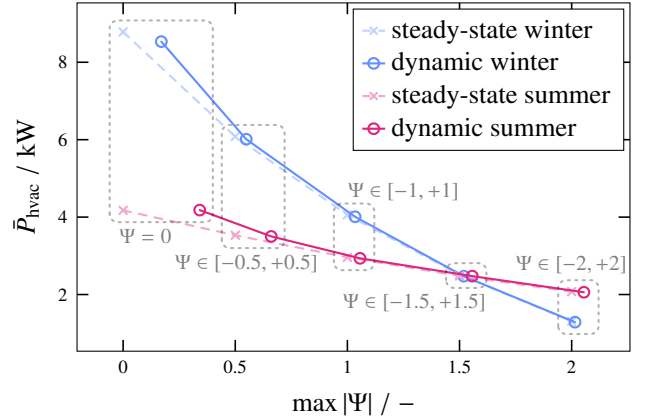

Analogous simulations for both the winter and summer missions are performed with five different \ac{PMV} constraints.
The resulting performance values are visualized in \cref{fig:dynamic-comparison-pareto}.
Each data point in \cref{fig:dynamic-comparison-pareto}, representing a pair of a dynamic and a static simulation, is a consolidation of entire simulation trajectories such as the ones shown in \cref{fig:dynamic-comparison-cold,fig:dynamic-comparison-hot}.
This consolidation is done based on the two primary performance indicators, i.e., the power consumption and the thermal comfort.
The first performance indicator is represented by the mean power consumption
\begin{equation}
\PhvacTimeAverage = \frac{1}{\tfinal - \tinit} \cdot \int_{\tinit}^{\tfinal} \Phvac(t^\prime) \, \mathrm{d} t^\prime ,
\label{eq:power-time-average}
\end{equation}
where \tinit and \tfinal represent the mission's initial and final time, respectively.
For the second performance indicator, we quantify the achieved \ac{PMV} window by evaluating the maximum absolute value of the \ac{PMV} trajectory.
For the dynamic trajectories, we first smooth the \ac{PMV} trajectory with a \qty{15}{\minute} moving average filter, which corresponds to a typical dwell time of passengers in a city bus.
Thus, this smoothed value represents an average comfort level experienced by passengers throughout a single ride.
The maximum absolute value of this smoothed trajectory thus represents the single most uncomfortable trip of any passenger during the entire day.
For this reason, we only consider \ac{PMV} values when at least one passenger is present in the bus.

\Cref{fig:dynamic-comparison-pareto} shows that the steady-state models are able to predict the behavior of the dynamic model with reasonable accuracy.
The larger deviations in the winter mission for high comfort requirements can be attributed to the power limit of the \ac{HP} of \qty{12}{\kilo\watt}.
This limits the ability of the controller to compensate for the effect of prolonged door openings.
As the door openings are averaged for the steady-state approach, such a temporary saturation does not influence the steady-state solution.
However, as shown in the next subsection, temperatures below \qty{0}{\celsius} are relatively rare in Zürich.
Thus, we do not expect this effect to significantly impact the annual performance metrics.
The larger deviations in the summer mission can be attributed to the delayed mode switch, as observed in \cref{fig:dynamic-comparison-hot} and explained above.

\begin{figure}
\centering
\definecolor{mycolor1}{rgb}{0.59375,1.00000,0.40625}%
\definecolor{mycolor2}{rgb}{0.39062,1.00000,0.60938}%
\definecolor{mycolor3}{rgb}{0.28125,1.00000,0.71875}%
\definecolor{mycolor4}{rgb}{0.06250,1.00000,0.93750}%
\definecolor{mycolor5}{rgb}{0.00000,0.42188,1.00000}%
\definecolor{mycolor6}{rgb}{0.00000,0.25000,1.00000}%
\definecolor{mycolor7}{rgb}{0.82353,0.82353,0.82353}%
\begin{tikzpicture}

\tikzset{every node/.style={
	font = {\footnotesize},
	inner sep = 2pt,
}}

\begin{axis}[%
width=0.97\linewidth,
scale only axis, %
axis equal image,
axis line style={draw=none},
ticks=none,
colormap/jet,
xmajorgrids=false,
ymajorgrids=false,
enlarge x limits = 0.005,
enlarge y limits = false,ymin=-0.35, %
]

\addplot[area legend, draw=none, fill=mycolor1, forget plot]
table[] {img/single-scenario-cold/single-scenario-cold-1-box-fill--Tcab.tsv}--cycle;
\node[fill=white, opacity = 0.6, text opacity = 1, align=center]
at (axis cs:0.75,1) {\Tcab\\\qty{16.1}{\celsius}};

\addplot[area legend, draw=black, forget plot]
table[] {img/single-scenario-cold/single-scenario-cold-2-box-edge--Tcab.tsv}--cycle;

\addplot[area legend, draw=none, fill=mycolor2, forget plot]
table[] {img/single-scenario-cold/single-scenario-cold-3-box-fill--Tint.tsv}--cycle;
\node[fill=white, opacity = 0.6, text opacity = 1, align=center]
at (axis cs:1,3.5) {\Tint\\\qty{13.5}{\celsius}};

\addplot[area legend, draw=black, forget plot]
table[] {img/single-scenario-cold/single-scenario-cold-4-box-edge--Tint.tsv}--cycle;

\addplot[area legend, draw=none, fill=mycolor3, forget plot]
table[] {img/single-scenario-cold/single-scenario-cold-5-box-fill--TRH.tsv}--cycle;
\node[fill=white, opacity = 0.6, text opacity = 1, align=center]
at (axis cs:0.75,5.5) {\TRH\\\qty{12.3}{\celsius}};

\addplot[area legend, draw=black, forget plot]
table[] {img/single-scenario-cold/single-scenario-cold-6-box-edge--TRH.tsv}--cycle;

\addplot[area legend, draw=none, fill=mycolor4, forget plot]
table[] {img/single-scenario-cold/single-scenario-cold-7-box-fill--TShellInside.tsv}--cycle;
\node[fill=white, opacity = 0.6, text opacity = 1, align=center]
at (axis cs:3,3.5) {\TShellInside\\\qty{9.5}{\celsius}};

\addplot[area legend, draw=black, forget plot]
table[] {img/single-scenario-cold/single-scenario-cold-8-box-edge--TShellInside.tsv}--cycle;

\addplot[area legend, draw=none, fill=mycolor5, forget plot]
table[] {img/single-scenario-cold/single-scenario-cold-9-box-fill--TShellOutside.tsv}--cycle;
\node[fill=white, opacity = 0.6, text opacity = 1, align=center]
at (axis cs:5,3.5) {\TShellOutside\\\qty{1.4}{\celsius}};

\addplot[area legend, draw=black, forget plot]
table[] {img/single-scenario-cold/single-scenario-cold-10-box-edge--TShellOutside.tsv}--cycle;

\addplot[area legend, draw=none, fill=mycolor6, forget plot]
table[] {img/single-scenario-cold/single-scenario-cold-11-box-fill--Tamb.tsv}--cycle;
\node[fill=white, opacity = 0.6, text opacity = 1, align=center]
at (axis cs:7,3) {\Tamb\\\qty{-0.7}{\celsius}};

\addplot[area legend, draw=black, forget plot]
table[] {img/single-scenario-cold/single-scenario-cold-12-box-edge--Tamb.tsv}--cycle;

\addplot[area legend, draw=black, fill=mycolor7, forget plot]
table[] {img/single-scenario-cold/single-scenario-cold-13-arrow--Qpass.tsv}--cycle;
\node[align=center] (QpassLbl)
at (axis cs:2.2,-0.05) {\Qpass\\\qty{4.2}{\kilo\watt}};

\addplot[area legend, draw=black, fill=mycolor7, forget plot]
table[] {img/single-scenario-cold/single-scenario-cold-14-arrow--Qother.tsv}--cycle;
\node[align=center]
at (axis cs:-0.5,-0.05) {\Qother\\\qty{1.0}{\kilo\watt}};

\addplot[area legend, draw=black, fill=mycolor7, forget plot]
table[] {img/single-scenario-cold/single-scenario-cold-15-arrow--Qdoor.tsv}--cycle;
\node[align=center]
at (axis cs:4,0.9) {\Qdoor\\\qty{2.3}{\kilo\watt}};

\addplot[area legend, draw=black, fill=mycolor7, forget plot]
table[] {img/single-scenario-cold/single-scenario-cold-16-arrow--QconvInt.tsv}--cycle;
\node[align=center, rotate=90]
at (axis cs:0.95,2.5) {\QconvInt\\\qty{0.4}{\kilo\watt}};

\addplot[area legend, draw=black, fill=mycolor7, forget plot]
table[] {img/single-scenario-cold/single-scenario-cold-17-arrow--QconvRH.tsv}--cycle;
\node[align=center, rotate=90]
at (axis cs:0.0,4.0) {\QconvRH\\\qty{0.0}{\kilo\watt}};

\addplot[area legend, draw=black, fill=mycolor7, forget plot]
table[] {img/single-scenario-cold/single-scenario-cold-18-arrow--QradRhToInt.tsv}--cycle;
\node[align=center, rotate=90]
at (axis cs:0.95,4.5) {\QradRhToInt\\\qty{0.0}{\kilo\watt}};

\addplot[area legend, draw=black, fill=mycolor7, forget plot]
table[] {img/single-scenario-cold/single-scenario-cold-19-arrow--QradRhToShell.tsv}--cycle;
\node[align=center]
at (axis cs:2,5.9) {\QradRhToShell\\\qty{0.0}{\kilo\watt}};

\addplot[area legend, draw=black, fill=mycolor7, forget plot]
table[] {img/single-scenario-cold/single-scenario-cold-20-arrow--QradIntToShell.tsv}--cycle;
\node[align=center]
at (axis cs:2,3.15) {\QradIntToShell\\\qty{0.4}{\kilo\watt}};

\addplot[area legend, draw=black, fill=mycolor7, forget plot]
table[] {img/single-scenario-cold/single-scenario-cold-21-arrow--QconvShellInside.tsv}--cycle;
\node[align=center]
at (axis cs:2,2.3) {\QconvShellInside\\\qty{10.6}{\kilo\watt}};

\addplot[area legend, draw=black, fill=mycolor7, forget plot]
table[] {img/single-scenario-cold/single-scenario-cold-22-arrow--Qcond.tsv}--cycle;
\node[align=center]
at (axis cs:4,4.4) {\Qcond\\\qty{11.0}{\kilo\watt}};

\addplot[area legend, draw=black, fill=mycolor7, forget plot]
table[] {img/single-scenario-cold/single-scenario-cold-23-arrow--QconvShellOutside.tsv}--cycle;
\node[align=center]
at (axis cs:6,4.1) {\QconvShellOutside\\\qty{9.0}{\kilo\watt}};

\addplot[area legend, draw=black, fill=mycolor7, forget plot]
table[] {img/single-scenario-cold/single-scenario-cold-24-arrow--QradShellOutside.tsv}--cycle;
\node[align=center]
at (axis cs:6,1.85) {\QradShellOutside\\\qty{2.0}{\kilo\watt}};

\addplot[area legend, draw=black, fill=mycolor7, forget plot]
table[] {img/single-scenario-cold/single-scenario-cold-25-arrow--Qhc.tsv}--cycle;
\node[align=center]
at (axis cs:-0.5,2.2) {\Qhc\\\qty{8.0}{\kilo\watt}};

\addplot[area legend, draw=black, fill=mycolor7, forget plot]
table[] {img/single-scenario-cold/single-scenario-cold-26-arrow--PRH.tsv}--cycle;
\node[align=center]
at (axis cs:-0.5,5.9) {\PRH\\\qty{0.0}{\kilo\watt}};

\addplot[area legend, draw=black, fill=mycolor7, forget plot]
table[] {img/single-scenario-cold/single-scenario-cold-27-arrow--QsolarInt.tsv}--cycle;
\node[align=center, rotate=-315]
at (axis cs:1.804,4.354) {\QsolarInt\\\qty{0.0}{\kilo\watt}};

\addplot[area legend, draw=black, fill=mycolor7, forget plot]
table[] {img/single-scenario-cold/single-scenario-cold-28-arrow--QsolarShellInside.tsv}--cycle;
\node[align=center, rotate=-315]
at (axis cs:3.804,6.354) {\QsolarShellInside\\\qty{0.0}{\kilo\watt}};

\addplot[area legend, draw=black, fill=mycolor7, forget plot]
table[] {img/single-scenario-cold/single-scenario-cold-29-arrow--QsolarShellOutside.tsv}--cycle;
\node[align=center, rotate=-315]
at (axis cs:5.804,6.354) {\QsolarShellOutside\\\qty{0.0}{\kilo\watt}};

\draw[->] (QpassLbl) to[bend left] (-0.5,0.8);

\end{axis}
\end{tikzpicture}%
\caption{
Visualization of the steady-state result corresponding to \DTMdisplay{2022}{12}{10}{-1}{17}{35}{0}{-1}{-1} to \DTMdisplaytime{18}{35}{0}, i.e., the shaded area in \cref{fig:dynamic-comparison-cold}.
In this sample, a \ac{PMV} of $\PMVvar=-1$ is reached.
The arrow annotations represent the corresponding absolute values of the flows.
}
\label{fig:single-scenario-cold}
\end{figure}
\begin{figure}
\centering
\definecolor{mycolor1}{rgb}{1.00000,0.70312,0.00000}%
\definecolor{mycolor2}{rgb}{1.00000,0.23438,0.00000}%
\definecolor{mycolor3}{rgb}{1.00000,0.35938,0.00000}%
\definecolor{mycolor4}{rgb}{0.90625,0.00000,0.00000}%
\definecolor{mycolor5}{rgb}{0.82353,0.82353,0.82353}%
\begin{tikzpicture}

\tikzset{every node/.style={
	font = {\footnotesize},
	inner sep = 2pt,
}}

\begin{axis}[%
width=0.97\linewidth,
scale only axis, %
axis equal image,
axis line style={draw=none},
ticks=none,
colormap/jet,
xmajorgrids=false,
ymajorgrids=false,
enlarge x limits = 0.005,
enlarge y limits = false,ymin=-0.35, %
]

\addplot[area legend, draw=none, fill=mycolor1, forget plot]
table[] {img/single-scenario-hot/single-scenario-hot-1-box-fill--Tcab.tsv}--cycle;
\node[fill=white, opacity = 0.6, text opacity = 1, align=center]
at (axis cs:0.75,1) {\Tcab\\\qty{24.9}{\celsius}};

\addplot[area legend, draw=black, forget plot]
table[] {img/single-scenario-hot/single-scenario-hot-2-box-edge--Tcab.tsv}--cycle;

\addplot[area legend, draw=none, fill=mycolor2, forget plot]
table[] {img/single-scenario-hot/single-scenario-hot-3-box-fill--Tint.tsv}--cycle;
\node[fill=white, opacity = 0.6, text opacity = 1, align=center]
at (axis cs:1,3.5) {\Tint\\\qty{30.8}{\celsius}};

\addplot[area legend, draw=black, forget plot]
table[] {img/single-scenario-hot/single-scenario-hot-4-box-edge--Tint.tsv}--cycle;

\addplot[area legend, draw=none, fill=mycolor3, forget plot]
table[] {img/single-scenario-hot/single-scenario-hot-5-box-fill--TRH.tsv}--cycle;
\node[fill=white, opacity = 0.6, text opacity = 1, align=center]
at (axis cs:0.75,5.5) {\TRH\\\qty{29.1}{\celsius}};

\addplot[area legend, draw=black, forget plot]
table[] {img/single-scenario-hot/single-scenario-hot-6-box-edge--TRH.tsv}--cycle;

\addplot[area legend, draw=none, fill=mycolor2!75!mycolor3, forget plot]
table[] {img/single-scenario-hot/single-scenario-hot-7-box-fill--TShellInside.tsv}--cycle;
\node[fill=white, opacity = 0.6, text opacity = 1, align=center]
at (axis cs:3,3.5) {\TShellInside\\\qty{30.3}{\celsius}};

\addplot[area legend, draw=black, forget plot]
table[] {img/single-scenario-hot/single-scenario-hot-8-box-edge--TShellInside.tsv}--cycle;

\addplot[area legend, draw=none, fill=mycolor4, forget plot]
table[] {img/single-scenario-hot/single-scenario-hot-9-box-fill--TShellOutside.tsv}--cycle;
\node[fill=white, opacity = 0.6, text opacity = 1, align=center]
at (axis cs:5,3.5) {\TShellOutside\\\qty{34.9}{\celsius}};

\addplot[area legend, draw=black, forget plot]
table[] {img/single-scenario-hot/single-scenario-hot-10-box-edge--TShellOutside.tsv}--cycle;

\addplot[area legend, draw=none, fill=mycolor4, forget plot]
table[] {img/single-scenario-hot/single-scenario-hot-11-box-fill--Tamb.tsv}--cycle;
\node[fill=white, opacity = 0.6, text opacity = 1, align=center]
at (axis cs:7,3) {\Tamb\\\qty{34.7}{\celsius}};

\addplot[area legend, draw=black, forget plot]
table[] {img/single-scenario-hot/single-scenario-hot-12-box-edge--Tamb.tsv}--cycle;

\addplot[area legend, draw=black, fill=mycolor5, forget plot]
table[] {img/single-scenario-hot/single-scenario-hot-13-arrow--Qpass.tsv}--cycle;
\node[align=center] (QpassLbl)
at (axis cs:2.2,-0.05) {\Qpass\\\qty{2.0}{\kilo\watt}};

\addplot[area legend, draw=black, fill=mycolor5, forget plot]
table[] {img/single-scenario-hot/single-scenario-hot-14-arrow--Qother.tsv}--cycle;
\node[align=center]
at (axis cs:-0.5,-0.05) {\Qother\\\qty{1.0}{\kilo\watt}};

\addplot[area legend, draw=black, fill=mycolor5, forget plot]
table[] {img/single-scenario-hot/single-scenario-hot-15-arrow--Qdoor.tsv}--cycle;
\node[align=center]
at (axis cs:4,0.9) {\Qdoor\\\qty{0.9}{\kilo\watt}};

\addplot[area legend, draw=black, fill=mycolor5, forget plot]
table[] {img/single-scenario-hot/single-scenario-hot-16-arrow--QconvInt.tsv}--cycle;
\node[align=center, rotate=-90]
at (axis cs:1.05,2.5) {\QconvInt\\\qty{0.9}{\kilo\watt}};

\addplot[area legend, draw=black, fill=mycolor5, forget plot]
table[] {img/single-scenario-hot/single-scenario-hot-17-arrow--QconvRH.tsv}--cycle;
\node[align=center, rotate=-90]
at (axis cs:0.05,4.0) {\QconvRH\\\qty{0.0}{\kilo\watt}};

\addplot[area legend, draw=black, fill=mycolor5, forget plot]
table[] {img/single-scenario-hot/single-scenario-hot-18-arrow--QradRhToInt.tsv}--cycle;
\node[align=center, rotate=90]
at (axis cs:0.95,4.5) {\QradRhToInt\\\qty{0.0}{\kilo\watt}};

\addplot[area legend, draw=black, fill=mycolor5, forget plot]
table[] {img/single-scenario-hot/single-scenario-hot-19-arrow--QradRhToShell.tsv}--cycle;
\node[align=center]
at (axis cs:2,5.9) {\QradRhToShell\\\qty{0.0}{\kilo\watt}};

\addplot[area legend, draw=black, fill=mycolor5, forget plot]
table[] {img/single-scenario-hot/single-scenario-hot-20-arrow--QradIntToShell.tsv}--cycle;
\node[align=center]
at (axis cs:2,3.15) {\QradIntToShell\\\qty{0.1}{\kilo\watt}};

\addplot[area legend, draw=black, fill=mycolor5, forget plot]
table[] {img/single-scenario-hot/single-scenario-hot-21-arrow--QconvShellInside.tsv}--cycle;
\node[align=center]
at (axis cs:2,2.25) {\QconvShellInside\\\qty{8.7}{\kilo\watt}};

\addplot[area legend, draw=black, fill=mycolor5, forget plot]
table[] {img/single-scenario-hot/single-scenario-hot-22-arrow--Qcond.tsv}--cycle;
\node[align=center]
at (axis cs:4,2.9) {\Qcond\\\qty{6.2}{\kilo\watt}};

\addplot[area legend, draw=black, fill=mycolor5, forget plot]
table[] {img/single-scenario-hot/single-scenario-hot-23-arrow--QconvShellOutside.tsv}--cycle;
\node[align=center]
at (axis cs:6,4.35) {\QconvShellOutside\\\qty{0.7}{\kilo\watt}};

\addplot[area legend, draw=black, fill=mycolor5, forget plot]
table[] {img/single-scenario-hot/single-scenario-hot-24-arrow--QradShellOutside.tsv}--cycle;
\node[align=center]
at (axis cs:6,1.85) {\QradShellOutside\\\qty{0.2}{\kilo\watt}};

\addplot[area legend, draw=black, fill=mycolor5, forget plot]
table[] {img/single-scenario-hot/single-scenario-hot-25-arrow--Qhc.tsv}--cycle;
\node[align=center]
at (axis cs:-0.4,2.45) {\Qhc\\\qty{13.5}{\kilo\watt}};

\addplot[area legend, draw=black, fill=mycolor5, forget plot]
table[] {img/single-scenario-hot/single-scenario-hot-26-arrow--PRH.tsv}--cycle;
\node[align=center]
at (axis cs:-0.5,5.9) {\PRH\\\qty{0.0}{\kilo\watt}};

\addplot[area legend, draw=black, fill=mycolor5, forget plot]
table[] {img/single-scenario-hot/single-scenario-hot-27-arrow--QsolarInt.tsv}--cycle;
\node[align=center, rotate=-315]
at (axis cs:1.804,4.354) {\QsolarInt\\\qty{1.0}{\kilo\watt}};

\addplot[area legend, draw=black, fill=mycolor5, forget plot]
table[] {img/single-scenario-hot/single-scenario-hot-28-arrow--QsolarShellInside.tsv}--cycle;
\node[align=center, rotate=-315]
at (axis cs:4.404,6.354) {\QsolarShellInside\\\qty{2.4}{\kilo\watt}};

\addplot[area legend, draw=black, fill=mycolor5, forget plot]
table[] {img/single-scenario-hot/single-scenario-hot-29-arrow--QsolarShellOutside.tsv}--cycle;
\node[align=center, rotate=-315]
at (axis cs:6.674,6.464) {\QsolarShellOutside\\\qty{7.1}{\kilo\watt}};

\draw[->] (QpassLbl) to[bend left] (-0.6,0.8);

\end{axis}
\end{tikzpicture}%
\caption{
Visualization of the steady-state result corresponding to \DTMdisplay{2019}{7}{24}{-1}{15}{50}{0}{-1}{-1} to \DTMdisplaytime{16}{50}{0}, i.e., the shaded area in \cref{fig:dynamic-comparison-hot}.
In this sample, a \ac{PMV} of $\PMVvar = 0.5$ is reached.
The arrow annotations represent the corresponding absolute values of the flows.
}
\label{fig:single-scenario-hot}
\end{figure}

\Cref{fig:single-scenario-cold,fig:single-scenario-hot} show example steady-state solutions from the winter and summer missions, respectively.
These visualizations serve to provide some intuition on the range of temperature levels and heat flows that are present in the model.

\subsection{Case Study 1: Offline Year-Round Evaluation}

In this case study, we show how the steady-state optimization approach can be used to evaluate the year-round performance of different \ac{HVAC} systems.
For this purpose, we first provide an overview of the entire set of samples that we use to represent the year-round operation of the \ac{HVAC} system.
We then explain how we post-process the results to obtain representative annual performance metrics.
Finally, we show the results of these calculations.

\begin{figure}
\centering
\begin{tikzpicture}

\begin{groupplot}[
	group style={
		group size=2 by 3,
        vertical sep=1.0cm,
        horizontal sep=0.9cm,
	},
    width=0.56\linewidth, 
    height=0.4\linewidth,
	ymin = 0,
    xlabel style={at={(0.5,-0.15)}, font=\small},
]

\nextgroupplot[
xlabel={\Tamb},
x unit = {\unit{\celsius}},
]
\addplot[ybar interval, fill=color1, fill opacity=0.6, draw=black, area legend] table[] {img/scenarios-overview/scenarios-overview-1-T--inf-.tsv};

\nextgroupplot[
xlabel={Irradiance},
x unit = {\unit{\watt\per\square\meter}},
legend style = {\legendNorthEast},
]
\addplot[ybar interval, fill=color1, fill opacity=0.6, draw=black, area legend] table[] {img/scenarios-overview/scenarios-overview-2-DNI.tsv};
\addlegendentry{\acs{DNI}}

\addplot[ybar interval, fill=color2, fill opacity=0.6, draw=black, area legend] table[] {img/scenarios-overview/scenarios-overview-3-DHI.tsv};
\addlegendentry{\acs{DHI}}

\nextgroupplot[
xlabel={\Npass},
x unit = {-},
]
\addplot[ybar interval, fill=color1, fill opacity=0.6, draw=black, area legend] table[] {img/scenarios-overview/scenarios-overview-4-N-P---.tsv};

\nextgroupplot[
xlabel={Month},
xtick = {1,3,5,7,9,11},
xticklabels = {Jan,Mar,May,Jul,Sep,Nov},
]
\addplot[ybar interval, fill=color1, fill opacity=0.6, draw=black, area legend] table[] {img/scenarios-overview/scenarios-overview-5-Months.tsv};

\nextgroupplot[
xlabel={\doorOpenFraction},
x unit = {\%},
]
\addplot[ybar interval, fill=color1, fill opacity=0.6, draw=black, area legend] table[] {img/scenarios-overview/scenarios-overview-6-x-zeta--door----.tsv};

\nextgroupplot[
xlabel={Time of day},
xticklabel={\pgfkeys{/pgf/number format/hour code={\tick}}},
xtick={0,6,12,18,24},
]
\addplot[ybar interval, fill=color1, fill opacity=0.6, draw=black, area legend] table[] {img/scenarios-overview/scenarios-overview-7-hours.tsv};
\end{groupplot}

\pgfresetboundingbox
\draw[use as bounding box,draw=none] (-0.8cm, -6.65cm) rectangle (7.7cm, 2cm);
\end{tikzpicture}%
\caption{Distributions of the values of the disturbances in the \glsmagnitude{Nsamples} samples.}
\label{fig:scenarios-overview}
\end{figure}
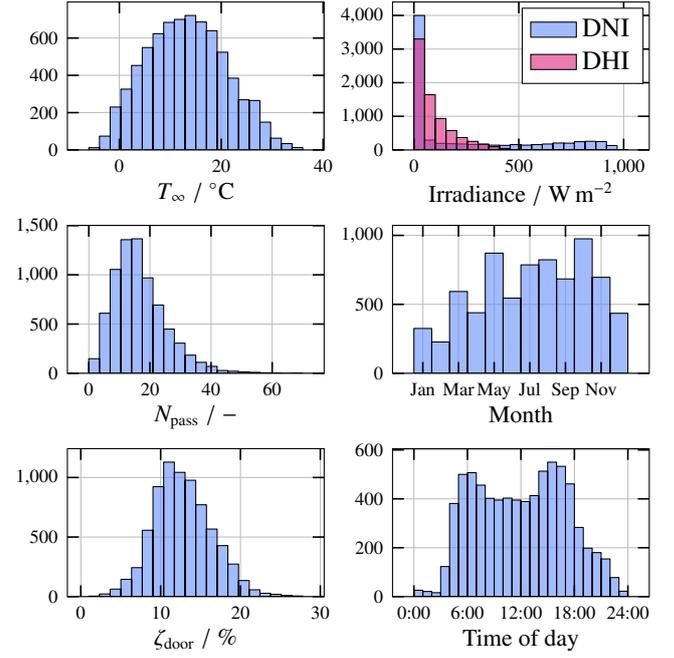

Based on the data sources introduced in \cref{sec:data-sources}, we create a dataset of constant disturbance values \disturbanceVector, consisting of \glsmagnitude{Nsamples} samples, each of which represents the average values in a one-hour segment.
\Cref{fig:scenarios-overview} provides an overview of this dataset.

As the samples are not uniformly distributed throughout the year, a simple time-average as introduced in \cref{eq:power-time-average} would not represent all seasons equally.
Thus, we calculate a weighted mean of the results.
Assuming sample $\idxi$ is recorded in January, for instance, we calculate its weight to be
\begin{equation}
\weight_\mathrm{,jan} = \frac{1}{325} \cdot \frac{31}{365} \,,
\end{equation}
where the first factor represents the 325 samples in January (see \cref{fig:scenarios-overview}) and the second factor represents the weight of January with its 31 days in relation to the entire year.
We use these weights to calculate the annual mean \ac{HVAC} system consumption:
\begin{equation}
\PhvacAnnualAverage = \sum_{\idxi=1}^{\Nsamples} (\weight \cdot \Phvac_{,\idxi}) .
\label{eq:meanPower}
\end{equation}

\begin{figure}
\centering
\begin{tikzpicture}

\begin{axis}[%
width=\linewidth,
height=0.6\linewidth,
xlabel={$\max \left| \PMVvar \right|$},
x unit = {-},
ylabel={\PhvacAnnualAverage},
y unit = kW,
xmin = 0,
xmax = 2,
ymin = 0,
ymax = 8,
legend style = {\legendNorthEast},
]

\addplot [color=color1, thick]
  table[]{img/pareto-fronts-thermal/pareto-fronts-thermal-1-PTC.tsv};
\addlegendentry{PTC}

\addplot [color=color2, thick]
  table[]{img/pareto-fronts-thermal/pareto-fronts-thermal-2-PTC-RH.tsv};
\addlegendentry{PTC + RH}

\addplot [color=color3, thick]
  table[]{img/pareto-fronts-thermal/pareto-fronts-thermal-3-HP.tsv};
\addlegendentry{HP}

\addplot [color=color4, thick]
  table[]{img/pareto-fronts-thermal/pareto-fronts-thermal-4-HP-AirCurtain.tsv};
\addlegendentry{HP + air curtain}

\end{axis}
\end{tikzpicture}%
\caption{Pareto front for different \ac{HVAC} system concepts showing the trade-off between the thermal comfort requirements and the mean annual power demand.
This analysis is based on all \glsmagnitude{Nsamples} samples.}
\label{fig:pareto-fronts-thermal}
\end{figure}
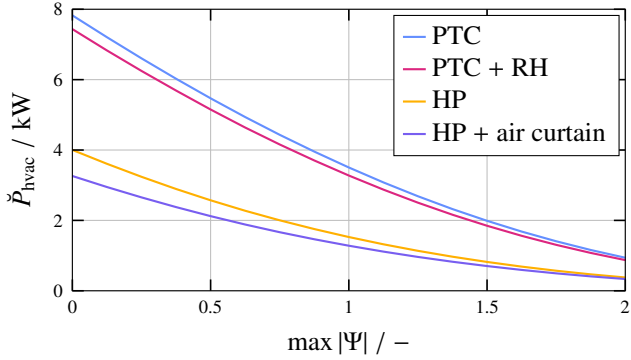

Such annual averages can be calculated for varying comfort requirements to obtain a Pareto front.
Such a Pareto front quantifies how much a certain comfort requirement costs in terms of mean power consumption.
Furthermore, it shows that increasing comfort levels becomes more and more expensive.
Comparing the Pareto fronts for various \ac{HVAC} designs, as shown in \cref{fig:pareto-fronts-thermal}, we observe that \acp{RH} can improve the mean energy efficiency of a \ac{PTC}-based heating system by 5 to 7\%.
While \pgls{HP} is a more expensive option, it also offers very significant benefits, with an annual reduction potential of 50 to 60\%.
Air curtains can further decrease the annual average consumption by 10 to 20\%.
As we have already shown in a previous study, a combination of \pgls{HP} with \acp{RH} does not yield significant additional benefits \cite{Widmer:2023df}.

\begin{figure}
\centering
\begin{tikzpicture}

\begin{axis}[%
width=\linewidth,
height=0.6\linewidth,
xlabel={\PMVvar},
x unit = -,
ylabel={Relative frequency},
y unit = -,
ymin = 0,
legend style = {\legendNorthEast},
]
\addplot[ybar interval, fill=pmv025Color, fill opacity=0.5, draw=black, area legend] table[] {img/yearly-pmv-range-comparison/yearly-pmv-range-comparison-1-PMV-0-25.tsv};
\addlegendentry{\PMVRange{0.25}}

\addplot[ybar interval, fill=color3, fill opacity=0.5, draw=black, area legend] table[] {img/yearly-pmv-range-comparison/yearly-pmv-range-comparison-2-PMV-0-62.tsv};
\addlegendentry{\PMVRange{0.625}}

\addplot[ybar interval, fill=pmv1Color, fill opacity=0.5, draw=black, area legend] table[] {img/yearly-pmv-range-comparison/yearly-pmv-range-comparison-3-PMV-1-00.tsv};
\addlegendentry{\PMVRange{1.0}}

\end{axis}
\end{tikzpicture}%
\caption{Overview of the optimized \ac{PMV} values in all \glsmagnitude{Nsamples} samples for varying comfort requirements.
The results are based on the \ac{HVAC} system with \pgls{HP} and air curtains.}
\label{fig:yearly-pmv-range-comparison}
\end{figure}
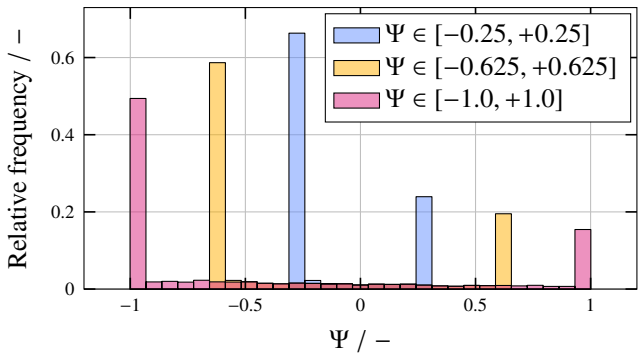

For the most efficient \ac{HVAC} system design, i.e., the \ac{HP} with air curtains, we visualize the resulting \ac{PMV} values in \cref{fig:yearly-pmv-range-comparison} for multiple comfort requirements.
Intuitively, whenever the constraint on the \ac{PMV} is active, the \ac{HVAC} system consumes power to satisfy this constraint.
For instance, \cref{fig:yearly-pmv-range-comparison} shows that the lower \ac{PMV} limit of $\PMVvar \ge -1$, is active in almost 50\% of the samples.
This means that heating is required almost 50\% of the time to achieve this comfort requirement.
Conversely, cooling is only required in less than 20\% of the samples.
This clearly demonstrates the fact that heating is the more relevant use case for the climatic conditions present in Zürich, despite the fact that January and February are slightly underrepresented in the samples considered in this case study.

\subsection{Case Study 2: Causal Online Control}
\label{sec:case-study-2-causal-online-control}

\newcommand{\GohlichEtAl}{Göhlich et al.\ \cite{Gohlich:2015}\xspace}
\newcommand{\HabraeusAndMinotta}{Hambraeus and Minotta \cite{Hambraeus:2023ej}\xspace}

In this section, we show how the optimization results based on the approach presented in \cref{sec:optimization} can be used in an online controller without the need for predictive data.
We do this by further focusing on the system with \pgls{HP} and air curtains.

\begin{figure*}[p]
\centering
\includegraphics{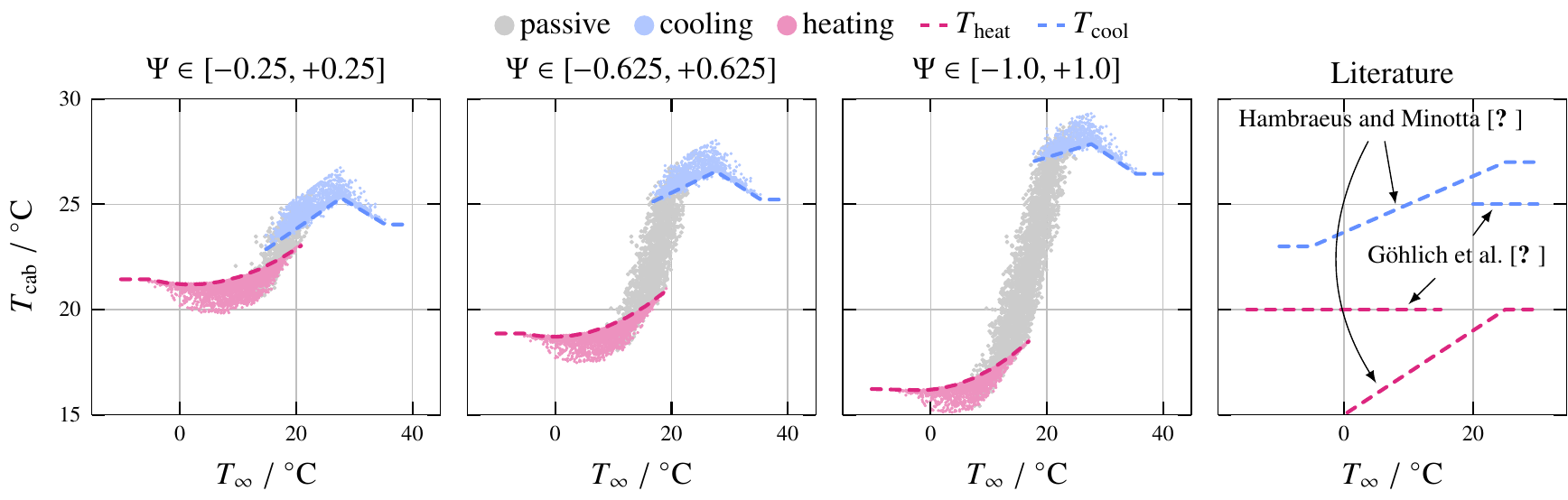}
\caption{Visualization of the extracted temperature setpoint profiles for three different comfort requirements (left three plots).
Each dot represents the result on one of the \glsmagnitude{Nsamples} samples.
The colors represent the active operating mode of the air heating and cooling unit.
The right plot shows two setpoint profiles from published literature.
}
\label{fig:temperature-setpoints}
\end{figure*}

Most \ac{HVAC} controllers used today track a cabin temperature setpoint, which in turn is typically dependent on the ambient temperature (see e.g. \cite{Hofstadter:2018mj,Dullinger:2015ub,Gohlich:2015,Hambraeus:2023ej}).
Our suggestion is thus to use the results from all \glsmagnitude{Nsamples} samples calculated in the previous section to generate such \enquote{setpoint profiles}, see \cref{fig:temperature-setpoints}.
Based on the shape of the point cloud, we use a quadratic fit for the heating profile and a piecewise linear fit for the cooling profile.
We select conservative fits for both modes (i.e., we fit the profiles to the upper edge of the point cloud for heating and to the lower edge for cooling) to maximize constraint satisfaction when the profiles are used.
These profiles can then be tracked by the \ac{PI} controller introduced in \cref{sec:ResultsValSS}.

\begin{figure*}[p]
\centering
\begin{minipage}[t]{0.48\textwidth}
\centering
\begin{tikzpicture}
\tikzset{every node/.style={
	font = {\small},
	inner sep = 2pt,
}}

\node[state, initial, align=center, anchor=center] at (0, 0) (passive) {$\operatingMode=0$\\\enquote{passive} };
\node[state, align=center, anchor=center] (heating) at (3.5, 1.5) {$\operatingMode=1$\\\enquote{heating} };
\node[state, align=center, anchor=center] (cooling) at (3.5, -1.5) {$\operatingMode=-1$\\\enquote{cooling} };

\draw
(passive) edge[->, bend left=20, above left, pos=0.9] node{$\Tcab < \Theat$} (heating)
(passive) edge[->, bend left=20, above right, pos=0.7] node{$\Tcab > \Tcool$} (cooling)
(passive) edge[->, loop above, min distance=0.75cm, left, pos=0.2] node{otherwise} (passive)
(heating) edge[->, loop right, min distance=0.75cm, above right, pos=0] node{otherwise} (heating)
(heating) edge[->, bend left=20, below right, pos=0.2] node{$\Tcab > \Theat + \qty{1}{\kelvin}$} (passive)
(cooling) edge[->, loop right, min distance=0.75cm, above right, pos=0] node{otherwise} (cooling)
(cooling) edge[->, bend left=20, below left, pos=0.3] node{$\Tcab < \Tcool - \qty{1}{\kelvin}$} (passive)
;
\end{tikzpicture}
\caption{State machine to determine the operating mode \operatingMode of the air heating and cooling unit.
Time dependencies are omitted to increase readability.
}
\label{fig:state-machine}
\end{minipage}
\hfill
\begin{minipage}[t]{0.48\textwidth}
\centering
\includegraphics{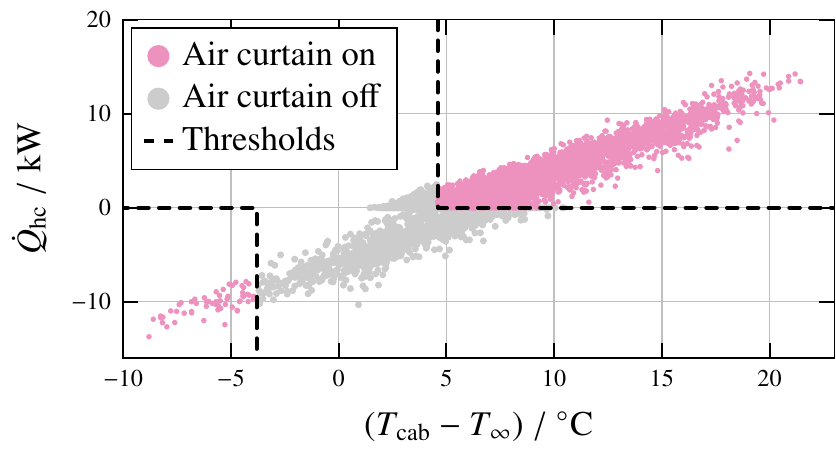}
\caption{Visualization of the air curtain usage patterns.
The black dashed lines visualize the two fitted thresholds for the heating mode ($\Qhc(t)>0$) and the cooling mode ($\Qhc(t)<0$), respectively.
This visualization is based on a comfort requirement of \PMVRange{1}, but the extracted thresholds are almost identical for different requirements.
}
\label{fig:aircurtain-setpoint}
\end{minipage}
\end{figure*}

The heating and cooling setpoint profiles can also be used to determine the operating mode $\operatingMode(t)$ of the air heating and cooling unit.
For this purpose, we propose a finite state machine, as illustrated in \cref{fig:state-machine}.
The constants of $\pm\qty{1}{\kelvin}$ in this state machine were tuned manually, such that frequent mode changes are prevented.

To determine the usage of the air curtain, we visualize the mode decision in all \glsmagnitude{Nsamples} samples in \cref{fig:aircurtain-setpoint}.
Based on this visualization, we determine a temperature difference threshold depending on the sign of $\Qhc(t)$ (or equivalently, $\operatingMode(t)$, according to \cref{eq:steady-state-hc}), which are also visualized in \cref{fig:aircurtain-setpoint}.
Since these temperature thresholds turn out to be almost independent of the selected comfort criteria, we use the same values throughout all subsequent analyses.

Next, we conduct a dynamic simulation based on the driving mission recorded on \DTMdisplaydate{2019}{07}{24}{-1}, i.e., the summer mission shown in \cref{fig:data-sample-hot,fig:dynamic-comparison-hot}, as this mission features highly dynamic ambient conditions.
We conduct multiple simulations to compare the temperature setpoints shown in \cref{fig:temperature-setpoints} with the corresponding setpoint profiles used in published literature, i.e., by \GohlichEtAl and by \HabraeusAndMinotta.
\Cref{fig:causal-time-resolved} shows the resulting trajectories of four dynamic simulations with different temperature setpoint profiles.
The setpoints are shown in the top plot, which qualitatively validates the respective tracking performance of the \ac{PI} controller.
The second and third plot show the achieved comfort values and the required total power, respectively.
While the heuristic setpoints from the literature achieve varying comfort values throughout the day, the setpoints extracted from the left plots of \cref{fig:temperature-setpoints} are able to maintain a more consistent comfort range over the dynamically changing ambient conditions.
\begin{figure*}[p]
\centering
\begin{tikzpicture}

\begin{groupplot}[
	group style={
		group size=1 by 3,
        vertical sep=0.2cm,
		xlabels at=edge bottom,
		xticklabels at=edge bottom,
	},
	width=\linewidth, 
    height=0.2\linewidth,
    xticklabel={\pgfkeys{/pgf/number format/hour code={\tick}}},
	y label style={ at={(-0.5cm,0.5)} },
	enlarge x limits = 0.01,
]

\nextgroupplot[
ylabel={\Tcab},
y unit = {\unit{\celsius}},
legend style={
	legend cell align=left, 
	at={(0.5, 1.0)}, 
	anchor=south,
    legend columns=-1,
	draw = none,
	fill=none
}
]
\addplot [pmv025Color!40!white, staticStyle]
  table[]{img/causal-time-resolved/causal-time-resolved-3-cabinStatic.tsv};

\addplot [pmv1Color!40!white, staticStyle]
  table[]{img/causal-time-resolved/causal-time-resolved-6-cabinStatic.tsv};

\addplot [pmv025Color, dynamicStyle]
  table[]{img/causal-time-resolved/causal-time-resolved-1-cabinDyn.tsv};
\addlegendentry{\PMVRange{0.25}}

\addplot [pmv025Color!40!white, refStyle, forget plot]
  table[]{img/causal-time-resolved/causal-time-resolved-2-cabinDynSet.tsv};

\addplot [pmv1Color, dynamicStyle]
  table[]{img/causal-time-resolved/causal-time-resolved-4-cabinDyn.tsv};
\addlegendentry{\PMVRange{1}}

\addplot [pmv1Color!40!white, refStyle, forget plot]
  table[]{img/causal-time-resolved/causal-time-resolved-5-cabinDynSet.tsv};

\addplot [gohlichColor, dynamicStyle]
  table[]{img/causal-time-resolved/causal-time-resolved-7-cabinDynG-hlich.tsv};
\addlegendentry{\GohlichEtAl}

\addplot [gohlichColor!40!white, refStyle, forget plot]
  table[]{img/causal-time-resolved/causal-time-resolved-8-cabinDynSet.tsv};

\addplot [hambraeusColor, dynamicStyle]
  table[]{img/causal-time-resolved/causal-time-resolved-9-cabinDynHam.tsv};
\addlegendentry{\HabraeusAndMinotta}

\addlegendimage{refStyle}
\addlegendentry{\Tcab setpoint}

\addplot [hambraeusColor!40!white, refStyle]
  table[]{img/causal-time-resolved/causal-time-resolved-10-cabinDynSet.tsv};

\nextgroupplot[
ylabel={\PMVvar},
y unit = {-},
]

\addplot [pmv025Color!40!white, staticStyle]
  table[]{img/causal-time-resolved/causal-time-resolved-14-PMVStatic.tsv};

\addplot [pmv1Color!40!white, staticStyle]
  table[]{img/causal-time-resolved/causal-time-resolved-16-PMVStatic.tsv};

\addplot [pmv025Color, dynamicStyle]
  table[]{img/causal-time-resolved/causal-time-resolved-13-PMVDyn.tsv};

\addplot [pmv1Color, dynamicStyle]
  table[]{img/causal-time-resolved/causal-time-resolved-15-PMVDyn.tsv};

\addplot [gohlichColor, dynamicStyle]
  table[]{img/causal-time-resolved/causal-time-resolved-17-PMVDynG-hlich.tsv};

\addplot [hambraeusColor, dynamicStyle]
  table[]{img/causal-time-resolved/causal-time-resolved-18-PMVDynHam.tsv};

\nextgroupplot[
ylabel={\Phvac},
y unit = kW,
]

\addplot [pmv025Color, dynamicStyle]
  table[]{img/causal-time-resolved/causal-time-resolved-20-P--hvac-Dyn.tsv};

\addplot [pmv1Color, dynamicStyle]
  table[]{img/causal-time-resolved/causal-time-resolved-22-P--hvac-Dyn.tsv};

\addplot [pmv025Color!40!white, staticStyle]
  table[]{img/causal-time-resolved/causal-time-resolved-21-P--hvac-Static.tsv};

\addplot [pmv1Color!40!white, staticStyle]
  table[]{img/causal-time-resolved/causal-time-resolved-23-P--hvac-Static.tsv};

\addplot [gohlichColor, dynamicStyle]
  table[]{img/causal-time-resolved/causal-time-resolved-24-PMVDynG-hlich.tsv};

\addplot [hambraeusColor, dynamicStyle]
  table[]{img/causal-time-resolved/causal-time-resolved-25-PMVDynHam.tsv};

\end{groupplot}
\end{tikzpicture}%
\caption{
Simulation results for \DTMdisplaydate{2019}{7}{24}{-1} based on the setpoint profiles shown in \cref{fig:temperature-setpoints,fig:aircurtain-setpoint}.
Thin solid lines represent the dynamic results, while thick dashed lines represent the results of the steady-state optimization.
The dotted black lines in the top plot represent the temperature setpoints if $\operatingMode(t) \in \{-1, 1\}$.
The temperature setpoint is not defined if no heating or cooling is used (i.e., $\operatingMode(t) = 0$).
}
\label{fig:causal-time-resolved}
\end{figure*}
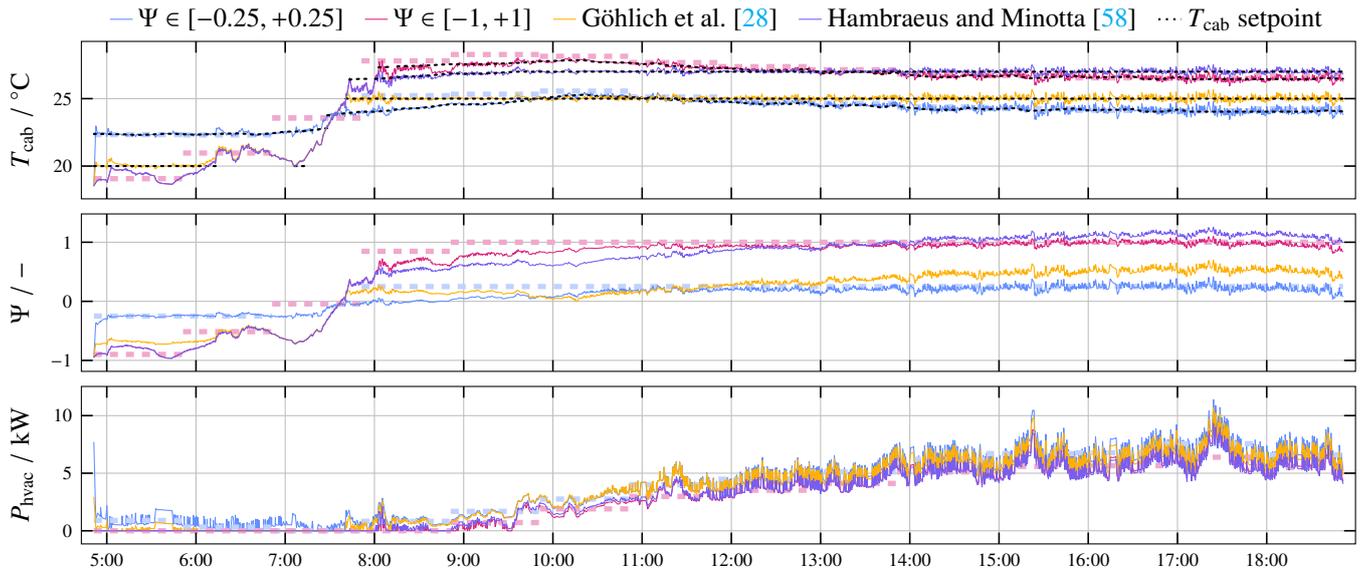

\begin{figure}
\centering
\begin{tikzpicture}

\pgfplotsset{
    pmvRegionStyle/.style={
		mark=o, 
        mark size=10,
		only marks,
		dash pattern=on 2pt off 2pt,
		thick,
	},
}

\begin{axis}[%
width=\linewidth,
height=0.7\linewidth,
xlabel={$\max \left| \PMVvar \right|$},
x unit = {-},
ylabel={\PhvacTimeAverage},
y unit = kW,
legend style={
    at={(0.45, 1.0)}, 
	anchor=south,
	legend columns=2,
	draw = none,
	font=\small, %
	fill=none
},
grid=none,
]
\addplot [black!40!white, 
dashed, 
mark=x, 
thick,
mark options={solid},
]
  table[]{img/causal-pareto-front/causal-pareto-front-1-static.tsv};
\addlegendentry{Steady-state optimization}

\addplot [black,
mark=o, 
thick,
]
  table[]{img/causal-pareto-front/causal-pareto-front-2-dynamic.tsv};
\addlegendentry{Extracted setpoints}

\addplot [gohlichColor, mark=*, mark size=3,
only marks,
mark options={solid},
]
table[]{img/causal-pareto-front/causal-pareto-front-3-G-hlich.tsv};
\addlegendentry{\GohlichEtAl}

\addplot [hambraeusColor, mark=*, mark size=3,
only marks,
mark options={solid},
]
table[]{img/causal-pareto-front/causal-pareto-front-4-Hambraeus.tsv};
\addlegendentry{\HabraeusAndMinotta}

\draw[pmvWindowMarkingRectangleStyle,pmv025Color,opacity=1] (axis cs:0.2,3.75) rectangle (axis cs:0.345,4.03);
\node[pmvWindowMarkingLabelStyle,anchor=west,color=pmv025Color] at (axis cs:0.345,3.95) {$\PMVRange{0.25}$};

\draw[pmvWindowMarkingRectangleStyle] (axis cs:0.575,3.28) rectangle (axis cs:0.71,3.52);
\node[pmvWindowMarkingLabelStyle,anchor=north east] at (axis cs:0.73,3.28) {$\PMVRange{0.625}$};

\draw[pmvWindowMarkingRectangleStyle,pmv1Color,opacity=1] (axis cs:0.95,2.86) rectangle (axis cs:1.073,3.09);
\node[pmvWindowMarkingLabelStyle,anchor=south west,color=pmv1Color] at (axis cs:0.97,3.07) {$\PMVRange{1}$};

\draw[pmvWindowMarkingRectangleStyle] (axis cs:1.45,2.38) rectangle (axis cs:1.56,2.63);
\node[pmvWindowMarkingLabelStyle,anchor=east] at (axis cs:1.45,2.45) {$\PMVRange{1.5}$};

\end{axis}
\end{tikzpicture}%
\caption{
Comparison of the results predicted with the steady-state approach and the results obtained with causal controllers based on different temperature setpoint profiles.
The dashed circles mark the solutions with the corresponding comfort requirements also shown in \cref{fig:causal-time-resolved}.
}
\label{fig:causal-pareto-front}
\end{figure}
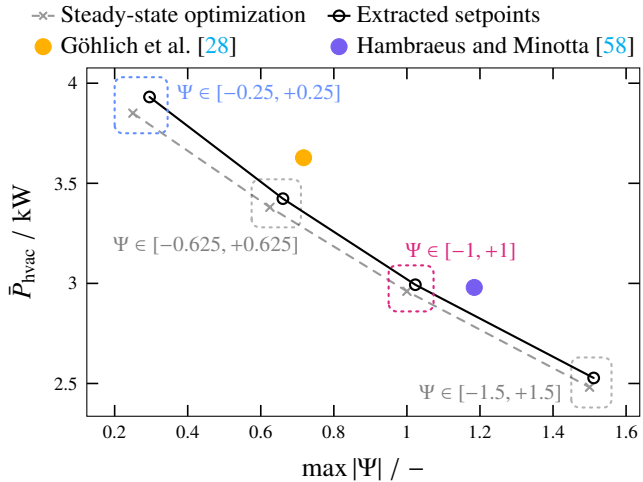

This observation is highlighted in \cref{fig:causal-pareto-front}, which again shows the maximum absolute value of the \ac{PMV} after being smoothed with a \qty{15}{\minute} moving average filter.
This figure shows that by extracting setpoints as shown in \cref{fig:temperature-setpoints,fig:aircurtain-setpoint} and using a causal controller, close-to-optimal performance is possible in terms of our performance indicators.
Compared to heuristic setpoint profiles from the literature, our approach reaches a better comfort value more consistently, resulting in a better trade-off as shown in \cref{fig:causal-pareto-front}.
Furthermore, the systematic approach to extract these setpoints allows for the bus operator to freely choose specific comfort targets.

\section{Conclusion}
\label{sec:conclusion}

\subsection{Contributions}

In this paper, we have presented a dynamic thermal model of the passenger cabin and the \ac{HVAC} system of a city bus, including a thermal comfort model. 
In a second step, we have introduced a steady-state approximation of this model.
We have then demonstrated both how the steady-state model enables a simplified formulation of a mathematical program to optimize the \ac{HVAC} system inputs and how it can be efficiently solved.
We have conducted a quantitative comparison between the dynamic and steady-state model.
While a steady-state approximation of the \ac{HVAC} system is not new in the context of public transportation vehicles (see e.g. \cite{Beusen:2013lg,Cvok:2021zt,Mansour:2018jz,Widmer:2023df,Wang:2021hy}), this comparative analysis is a novel contribution.
To showcase the practical usefulness of our methodology, we have conducted two case studies.
In the first case study, we have compared the annual performance of various \ac{HVAC} system configurations.
In the second case study, we have illustrated how our results can be utilized to derive optimized setpoints for causal controllers.

\subsection{Limitations and Outlook}

While our study provides valuable insights, we see certain limitations and opportunities for future research.
Although most of the presented model components are either validated or based on well-founded assumptions, a comprehensive validation of the entire thermal model is pending.
To address this, a measurement campaign is planned with a prototype vehicle, which will be used to ensure the accuracy of the models.

Future research could explore additional factors related to passenger comfort, such as humidity, air quality, ventilation-induced draft, radiant asymmetry, or heated seats.
Including these elements would further enrich the model and enable extended analyses of emerging trade-offs.
Extending the objective to include investment costs or embodied emissions enables holistic comparisons of \ac{HVAC} systems within frameworks like \ac{LCC} or \ac{LCA}.
Conducting case studies for different climatic conditions would allow to identify optimal \ac{HVAC} system configurations based on the deployment location.
Lastly, future studies could also extend to case studies encompassing multiple types of public transportation vehicles, such as street cars and trains.

\section*{CRediT Authorship Contribution Statement}

\textbf{Fabio Widmer:} Conceptualization, Methodology, Software, Validation, Visualization, Writing -- original draft
\textbf{Stijn van Dooren:} Conceptualization, Methodology, Validation, Visualization, Writing -- review \& editing
\textbf{Chris\-topher H.\ Onder:} Conceptualization, Funding acquisition, Project administration, Supervision

\section*{Declaration of Competing Interest}

The authors declare that they have no known competing financial interests or personal relationships that could have appeared to influence the work reported in this paper.

\section*{Acknowledgments}

This work was financially supported by the \acl{SFOE} (\acs{SFOE}, contract number SI/502235).
The industrial partners Carrosserie HESS AG and \ac{VBZ} provided assistance during data collection, model formulation, and validation.
None of the partners played a role in the writing of the article.

\bibliography{bibliography}

\appendix

\gdef\thesection{\Alph{section}} %
\makeatletter
\renewcommand\@seccntformat[1]{\appendixname\ \csname the#1\endcsname.\hspace{0.5em}}
\makeatother

\section{Model Constants}
\label{sec:model-constants}

\newglossarystyle{symbolDescriptionValue}{%
\setglossarystyle{long4col}%
\renewenvironment{theglossary}%
{\begin{supertabular}{ll@{\quad}r@{\,}l}}
{\end{supertabular}}
\renewcommand*{\glossaryheader}{}%
\renewcommand*{\glsgroupheading}[1]{}%
\renewcommand{\glossentry}[2]{%
\glstarget{##1}{\glossentryname{##1}} & \Glsentrydesc{##1} & \glsmagnitude{##1} & \glsunit{##1}\tabularnewline%
}
\renewcommand*{\glsgroupskip}{\relax}%
}

\begin{small}
\sisetup{per-mode=fraction} %
\renewcommand{\arraystretch}{1.1} %
\renewcommand{\glossarysection}[2][]{} %

\printnoidxglossary[
type=constants, %
sort=word, %
style=symbolDescriptionValue,
numberedsection=false, %
]
\end{small}

\end{document}